\documentclass[pdflatex,sn-mathphys-num]{sn-jnl}

\usepackage{graphicx}%
\usepackage{multirow}%
\usepackage{amsmath,amssymb,amsfonts}%
\usepackage{amsthm}%
\usepackage{mathrsfs}%
\usepackage[title]{appendix}%
\usepackage{xcolor}%
\usepackage{textcomp}%
\usepackage{manyfoot}%
\usepackage{booktabs}%
\usepackage{algorithm}%
\usepackage{algorithmicx}%
\usepackage{algpseudocode}%
\usepackage{listings}%
\usepackage{float}
\usepackage{comment}
\usepackage{mathtools}
\usepackage{etoolbox}   
\makeatletter
\patchcmd{\@maketitle}{\artauthors}{{\artauthors}}{}{}
\makeatother

\begin{document}

\title{Scalable Construction of Spiking Neural Networks using up to thousands of GPUs}


\author[1,2]{\fnm{Bruno} \sur{Golosio}}\equalcont{These authors contributed equally to this work.}

\author*[2,1]{\fnm{Gianmarco} \sur{Tiddia}}\email{gianmarco.tiddia@dsf.unica.it}\equalcont{These authors contributed equally to this work.}

\author[3,4]{\fnm{José} \sur{Villamar}}\equalcont{These authors contributed equally to this work.}

\author[5]{\fnm{Luca} \sur{Pontisso}}
\author[1,2]{\fnm{Luca} \sur{Sergi}}
\author[5]{\fnm{Francesco} \sur{Simula}}
\author[6,7]{\fnm{Pooja} \sur{Babu}}
\author[5]{\fnm{Elena} \sur{Pastorelli}}
\author[3,7]{\fnm{Abigail} \sur{Morrison}}
\author[3,8,9,10]{\fnm{Markus} \sur{Diesmann}}
\author[5]{\fnm{Alessandro} \sur{Lonardo}}
\author[5]{\fnm{Pier Stanislao} \sur{Paolucci}}
\author[3,11]{\fnm{Johanna} \sur{Senk}}

\affil[1]{\orgdiv{Department of Physics}, \orgname{University of Cagliari}, \orgaddress{\city{Monserrato}, \country{Italy}}}

\affil[2]{\orgdiv{Istituto Nazionale di Fisica Nucleare (INFN)}, \orgname{Sezione di Cagliari}, \orgaddress{\city{Monserrato}, \country{Italy}}}

\affil[3]{\orgdiv{Institute for Advanced Simulation (IAS-6)}, \orgname{Jülich Research Centre}, \orgaddress{\city{Jülich}, \country{Germany}}}

\affil[4]{\orgname{RWTH Aachen University}, \orgaddress{\city{Aachen}, \country{Germany}}}

\affil[5]{\orgdiv{Istituto Nazionale di Fisica Nucleare (INFN)}, \orgname{Sezione di Roma}, \orgaddress{\city{Rome}, \country{Italy}}}

\affil[6]{\orgdiv{Simulation and Data Lab Neuroscience, Jülich Supercomputer Centre}, \orgname {Jülich Research Centre}, \orgaddress{\city{Jülich}, \country{Germany}}}

\affil[7]{\orgdiv{Department of Computer Science 3 - Software Engineering}, \orgname{RWTH Aachen University}, \orgaddress{\city{Aachen}, \country{Germany}}}

\affil[8]{\orgdiv{JARA-Institute Brain Structure-Function Relationships (INM-10)}, \orgname{Jülich Research Centre}, \orgaddress{\city{Jülich}, \country{Germany}}}

\affil[9]{\orgdiv{Department of Psychiatry, Psychotherapy and Psychosomatics, School of Medicine}, \orgname{RWTH Aachen University}, \orgaddress{\city{Aachen}, \country{Germany}}}

\affil[10]{\orgdiv{Department of Physics, Faculty 1}, \orgname{RWTH Aachen University}, \orgaddress{\city{Aachen}, \country{Germany}}}

\affil[11]{\orgdiv{Sussex AI, School of Engineering and Informatics}, \orgname{University of Sussex}, \orgaddress{\city{Brighton}, \country{United Kingdom}}}

\abstract{Diverse scientific and engineering research areas deal with discrete, time-stamped changes in large systems of interacting delay differential equations. Simulating such complex systems at scale on high-performance computing clusters demands efficient management of communication and memory. Inspired by the human cerebral cortex — a sparsely connected network of $\mathcal{O}(10^{10})$ neurons, each forming $\mathcal{O}(10^{3})$--$\mathcal{O}(10^{4})$ synapses and communicating via short electrical pulses called spikes — we study the simulation of large-scale spiking neural networks for computational neuroscience research. This work presents a novel network construction method for multi-GPU clusters and upcoming exascale supercomputers using the Message Passing Interface (MPI), where each process builds its local connectivity and prepares the data structures for efficient spike exchange across the cluster during state propagation. We demonstrate scaling performance of two cortical models using point-to-point and collective communication, respectively.}


\maketitle

Each one of the billions of cortical neurons sends up to around ten action potentials (spikes) per second with sub-millisecond temporal precision to thousands of other neurons through a network with intricate, hierarchical connectivity structure. Therefore, each cortical microcircuit below $1$\,mm$^2$ of brain surface can produce one billion synaptic events per second of biological activity~\cite{Senk2026}. This rough estimate points to two main challenges posed by neuroscience research on simulation technology:
1)\,brain-scale simulations at the level of individual neurons and synapses require substantial amounts of computer memory,
2)\,resolving the neuronal and synaptic dynamics with sufficient accuracy even in long simulations requires high processing speeds~\cite{Aimone2023}.
High-Performance Computing (HPC) provides high-end hardware in pre-exascale and exascale supercomputers. Drawing their compute power from thousands of general-purpose GPUs connected with InfiniBand or other high-speed networks~\cite{Turisini2024}, these systems are promising candidates for approaching the required scales and speeds.
To harness these distributed systems effectively, simulation software with parallelized algorithms and efficient information exchange between compute nodes needs to be developed and optimized.
The Message Passing Interface (MPI)~\citep{MPIForum09} is the main standard for inter-node communication, offering both point-to-point and collective communication between groups of processes.

So far, no GPU-based simulation software for point-neuron spiking neural networks (SNNs) has been reported that scales up to entire compute clusters, fulfilling the infrastructure and accuracy demands of computational neuroscience. Alternative approaches that achieve a subset of these goals include the Digital Brain~\citep{Lu2024, Du2024}, which simulates networks with up to $10^{10}$ neurons on thousands of GPUs using MPI, but without storing individual synapse information.
Similarly, GeNN~\citep{Yavuz2016} simulates millions of neurons on a single GPU using an on-the-fly connectivity generation method~\cite{Knight2021_136}.
CARLsim supports multi-GPU execution on a single machine \cite{beyeler2015carlsim,CARLsim6}.
NEURON~\cite{Carnevale_2006}, enhanced by CoreNEURON, already combines MPI with GPU acceleration with a focus on multi-compartment neuron models~\cite{kumbhar2019coreneuron, Awile2022}. They demonstrate the parallel evaluation of a large number of differential equations required to update the complex neuron dynamics.
In contrast, the main bottleneck in distributed simulations of large networks of point neurons is the spike communication between different nodes of a compute cluster, and this holds for nodes equipped with GPUs or CPUs~\cite{Pronold2022, Tiddia2022}. STACS~\cite{Wang2015} tried to overcome this bottleneck by leveraging Charm++ parallelization capabilities, showing good scaling performance in the simulation of balanced network models~\cite{Wang2024}.

The simulation code NEST~\cite{Gewaltig_07_11204} employs a hybrid approach using MPI for inter-node and OpenMP~\citep{OpenMPSpec} for intra-node parallelization on modern multi-core processors (CPUs).
Benchmark SNN models have reached up to $10^9$ neurons and $10^{13}$ synapses~\cite{Kunkel2014, Jordan2018}, while models representing distinct brain regions with millions of neurons and billions of synapses have been simulated using thousands of cores in parallel~\cite{Schmidt2018, Pronold24_409, Senk24_405, Gandolfi2023, Yamaura2020}.
These efforts have made NEST a reference platform for large-scale SNN simulations~\cite{Senk2026}, keeping pace with conventional hardware development and the increasing interest in scalable neuromorphic systems~\cite{Kudithipudi2025}.
The round-robin algorithm to distribute neurons on MPI processes and the collective communication~\citep{Jordan2018} for synchronous spike exchange between processes used by NEST is optimal for load balancing assuming homogeneous network structure and neuron activity.
However, this approach does not exploit the heterogeneity of biological brains, including their topological and spatial multi-scale network organization and transient changes in spiking statistics~\citep{Albers2022}.

The recent GPU implementation of NEST addresses these demands.
NEST GPU~\cite{Golosio2021, Golosio23_9598} supports multi-GPU simulations and takes advantage of the locality of neuron populations. Like the CPU version, the C++ code exposes a handful of functions to the Python level for the setup and execution of concrete network models, with some of them being validated against the CPU version \cite{Golosio2021, Tiddia2022}. The code uses CUDA to interface GPU hardware and MPI for point-to-point communication between compute nodes~\cite{Tiddia2022}.
The initial algorithm for network construction relied on a transfer from CPU to GPU memory.
To reduce computation times, \cite{Golosio23_9598} proposed a method to dynamically build the network directly in GPU memory, applicable to single-GPU simulations.

Going beyond these previous works, here we present memory-efficient data structures for in-GPU network construction and spike communication during simulation on multi-GPU systems. Each MPI process builds its part of the network without MPI communication with other processes, creates the data structures needed to communicate via MPI, and efficiently transmits spikes across remote connections using either point-to-point or collective communication according to the user's choice.
As demonstrated by a reference implementation, the new method enables large-scale spiking neural network simulations of up to thousands of GPUs exploiting a computational communication scheme inspired by the temporal and spatial sparsity observed in the brain. Our results indicate that network sizes of $2\times 10^{10}$ neurons and $10^{14}$ synapses can be reached with our approach on the upcoming exascale supercomputer JUPITER~\cite{Herten24_1}.

\section*{Results}

Simulations of neuronal networks typically consist of two main phases: \textit{construction} and \textit{simulation} (or \textit{state propagation}). The former comprises all the steps needed before a loop advancing the dynamical state of the network begins. In our reference implementation, the network construction can be conceptually divided into subtasks such as the initialization phase, which includes the creation of neurons, devices (for stimulation or recording), and connections, and a phase that organizes data structures for spike delivery (here named \textit{simulation preparation}); see also Section \ref{sec:benchmarking} for details on the phases.

\begin{figure}[H]
    \centering
    \includegraphics[width=1\linewidth]{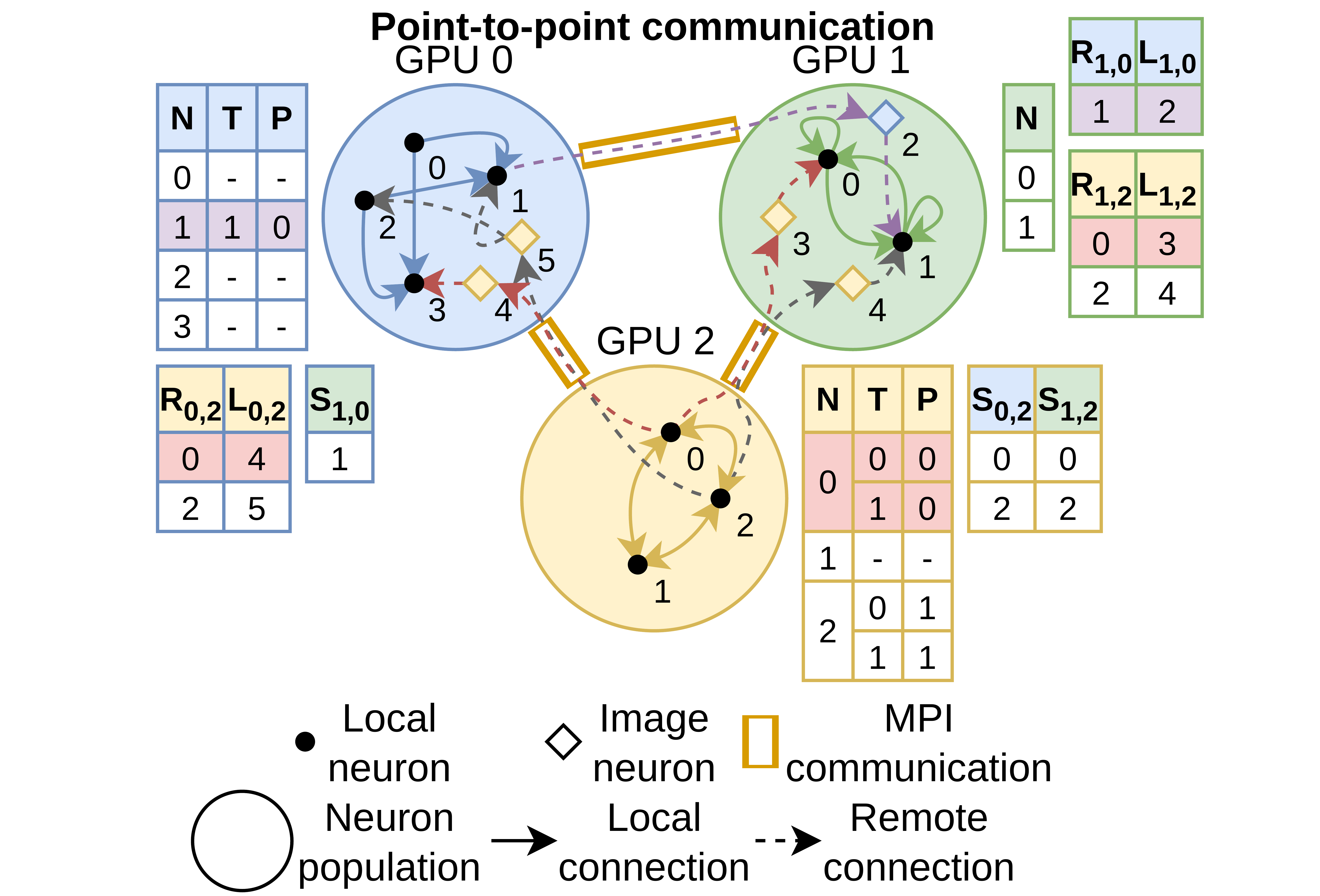}
    \caption{
Communication scheme for point-to-point spike routing and delivery with one MPI process per GPU.
Each process is identified by rank and color ($0$: blue, $1$: green, $2$: yellow).
After neuron creation, all ranks contain an arbitrary number of neurons, each with a unique index 
in the range from $0$ to $M_{\sigma} -1$, where $M_{\sigma}$ is the number of neurons in the rank $\sigma$.
The $(\mathbf{R}_{\tau, \sigma}, \mathbf{L}_{\tau, \sigma})$ maps associate the index of the source neuron in rank $\sigma$ to the index of its image in the target rank $\tau$.
The (\textbf{N}, \textbf{T}, \textbf{P}) tables associate to each local neuron, identified by the table row index \textbf{N}, the target processes \textbf{T} in which it has an image, and the corresponding position \textbf{P} of this image in the $(\mathbf{R}_{\tau, \sigma}, \mathbf{L}_{\tau, \sigma})$ maps.
Spikes emitted by neuron $1$ of rank $0$ and neurons $0$ and $2$ of rank $2$ are routed to their proxies in the corresponding target ranks, listed in array $\mathbf{T}$, by sending the positions $\mathbf{P}$ in the maps $(\mathbf{R}_{\tau, \sigma}, \mathbf{L}_{\tau, \sigma})$ through MPI.
$\mathbf{S}_{\tau, \sigma}$ in the source process corresponds to $\mathbf{R}_{\tau, \sigma}$ in the target process.
Within each rank, remote connections from image neurons serve as the final link to local neurons receiving spikes.
As an example (in red), when neuron $0$ of rank $2$ emits a spike, the (\textbf{N}, \textbf{T}, \textbf{P}) tables of that rank (bottom right, yellow heading) are used to retrieve the ranks of the target processes, \textbf{T}, in which this neuron has images ($0$ and $1$ in this example), and the positions \textbf{P} of those images ($0$ in both cases)
in the    
$(\mathbf{R}_{0, 2}, \mathbf{L}_{0, 2})$ and $(\mathbf{R}_{1, 2}, \mathbf{L}_{1, 2})$ maps, respectively.
Those positions are sent through MPI by point-to-point communication (orange) and they are used by the target MPI processes (ranks $0$ and $1$)
to retrieve the local indexes of the image nodes from the \textbf{L} columns of the maps. The spike is then treated as a local spike and is therefore sent to the target neurons through the outgoing connections of the image neuron.
Similarly (in purple), spikes emitted by neuron $1$ of rank $0$ are routed to the image neuron $2$ in rank $1$. It can be observed that rank $1$ has two (\textbf{R}, \textbf{L}) maps (top right, blue and yellow headings) because it has images of neurons from both $0$ and $2$ ranks.
    }
    \label{fig:p2p_routing}
\end{figure}

\begin{figure}[H]
    \centering
    \includegraphics[width=1\linewidth]{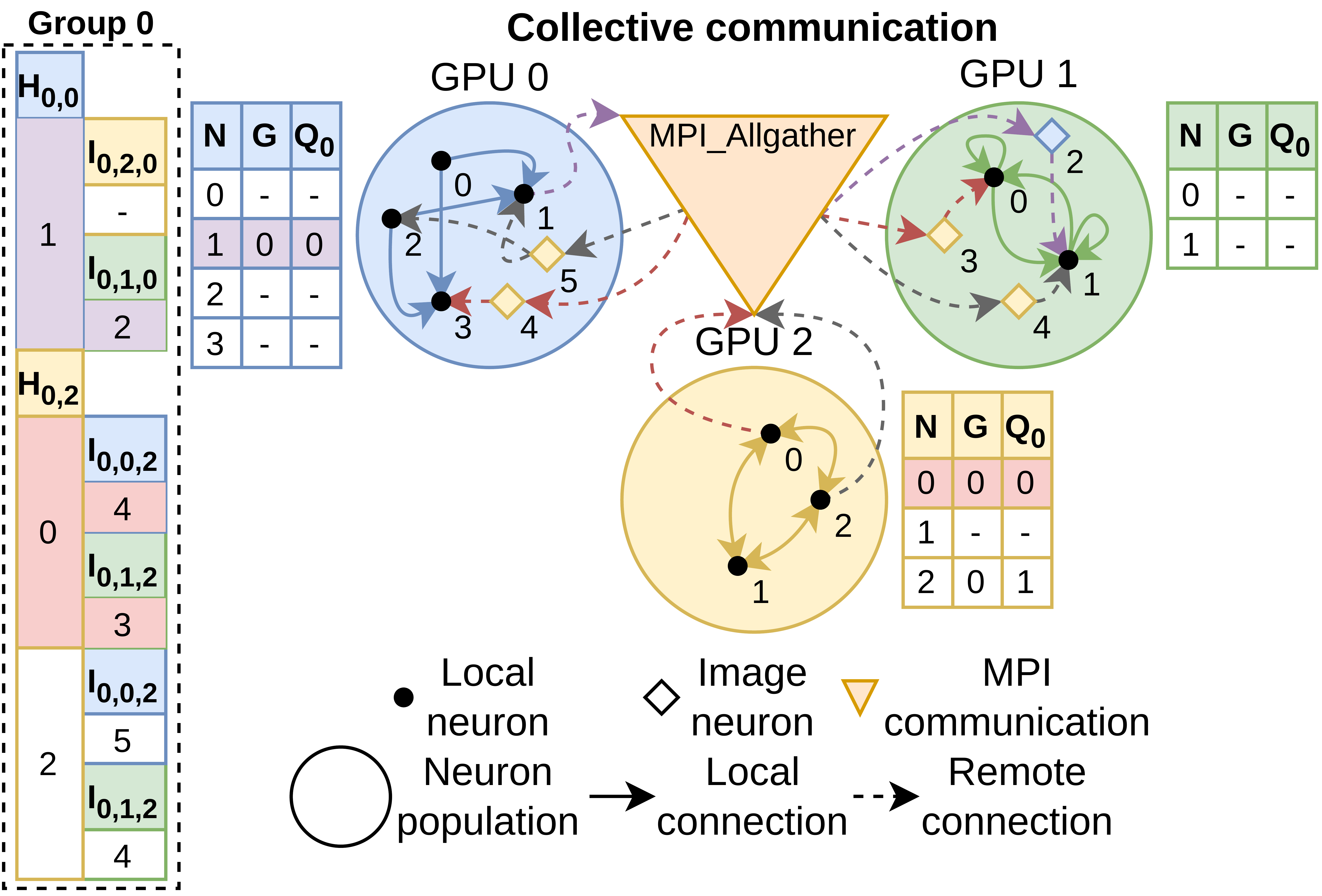}
    \caption{
Communication scheme for collective spike routing and delivery with one MPI process per GPU.
Each process is identified by rank and color ($0$: blue, $1$: green, $2$: yellow).
After neuron creation, all ranks contain an arbitrary number of neurons, each with a unique index in the rank local neuron array \textbf{N}.
All ranks belonging to an MPI group use specific indexing arrays mirrored across all members of the group, here group $0$ uses arrays denoted with suffix $0$.
Neurons of rank $\sigma$, which have a proxy in any other rank of the group, are indexed in the host array, here $\mathbf{H}_{0, \sigma}$.
Spikes emitted by neurons $1$ of rank $0$ and neurons $0$ and $2$ of rank $2$
are first routed to the groups to which they have remote connections using the group array $\mathbf{G}$, simultaneous participation to multiple groups is possible; in this illustration only group $0$ exits.
$\mathbf{Q}_0$ stores the positions of source neuron indexes in the host array, these are used to exchange the indexes of spiking neurons by all members of the group, and once received, spikes are routed to the image neurons using image index array $\mathbf{I}_{0, \tau, \sigma}$.
Within each rank, remote connections from image neurons serve as the final link to local neurons receiving spikes.
As an example (in red), when neuron $0$ of rank $2$ emits a spike, the (\textbf{N}, \textbf{G}, $\mathbf{Q}_0$) tables of that rank (bottom right, yellow heading) are used to retrieve the communication groups from \textbf{G} in which this neuron has images (only group $0$ this example), and its position in the host specific source index table $\mathbf{H}_{0, 2}$ from table $\mathbf{Q}_0$ ($0$ in this example).
Its position is sent through MPI by collective communication (orange) and is received by both ranks $0$ and $1$.
Rank $0$ uses the map $\mathbf{I}_{0, 0, 2}$ to retrieve the local index of the image node, and treats the incoming spike spike as a local spike, therefore sending the spike to the target neurons through the outgoing connections of the image neuron.
In the same way, rank $1$ uses the map $\mathbf{I}_{0, 1, 2}$ to retrieve the local index of the image node to route the spike to local targets.
Similarly (in purple), spikes emitted by neuron $1$ of rank $0$ are routed to the image neuron $2$ in rank $1$.
    }
    \label{fig:collective_routing}
\end{figure}

The main result of this work is the design of a novel algorithm for network construction that instantiates and organizes contiguous communication maps in GPU memory for efficient spike routing and delivery.
Similar to the parallel network construction methods previously developed by \citet{Jordan2018}, this algorithm relies on each MPI process independently building local connectivity and preparing data structures to manage spikes transmitted to neurons on other processes, avoiding communication between MPI processes during this phase.
Here, we distinguish between local connections, linking neurons within the same MPI process, and remote connections, spanning neurons across different processes.
Our method uses proxy representations of neurons for remote connections, which are an image of source neurons in the target processes. With each MPI process handling a single GPU, connections can then be generated locally in GPU memory following the approach described in~\cite{Golosio23_9598}.
Figures \ref{fig:p2p_routing} and \ref{fig:collective_routing} illustrate how our communication maps are used for spike routing and delivery and further details can be found in Section~\ref{sec:remote_conns}.
The mapping structures support both point-to-point and collective MPI communication, providing flexibility to adapt to network architecture and activity patterns. The following subsections elaborate on these two communication schemes and show performance results.

\subsection{Point-to-point communication: multi-area model} \label{sec:results_mam}
The Multi-Area Model (MAM) of $32$ vision-related areas of macaque monkey cortex~\cite{Schmidt2017, Schmidt2018} serves as a neuroscientifically relevant model to assess simulator performance in a network with complex, hierarchical architecture and biologically plausible spike statistics. The model represents each area by a $1$\,mm$^2$ patch with the full thickness of the cortex and natural density of neurons and synapses, comprising a total of $4.13 \times 10^6$ neurons and $24.2 \times 10^9$ synapses. See Section \ref{sec:MAM} for additional details on the model. Simulations are performed in the model’s \textit{metastable state}, reflecting resting-state activity in lightly anesthetized monkeys. We use the MPI point-to-point communication protocol for this model.
\textit{Point-to-point} protocols enable direct communication between two processes and are advantageous for networks with uneven distribution of neurons or synapses, resulting in highly heterogeneous communication patterns between processes.

We previously used this model to explore the potential of multi-GPU systems for speeding up the state propagation phase~\cite{Tiddia2022}. In that work, the network construction phase is carried out on the CPU. As in~\cite{Golosio23_9598}, we call this algorithm \textit{offboard}, to distinguish it from the new \textit{onboard} version that performs this phase directly on the GPU. Moreover, in~\cite{Tiddia2022} we compared the results of the spike statistics with the CPU version of the NEST simulator. Similarly to what has been done in that work, we here validated the results of the spike statistics of the model with respect to the previous version of the simulation code. Despite the high degree of variability in the MAM metastable state \cite{Schmidt2018}, the validation, introduced in Section \ref{sec:validation_spikes} and presented in detail in Appendix \ref{app:validation_spikes}, underlines the compatibility between the two versions.

Here, simulations follow the same configuration as~\cite{Tiddia2022} for the \textit{offboard} version, using $32$ NVIDIA V100 GPUs on the JUSUF cluster~\cite{VonStVieth2021}, with one GPU per cortical area due to the memory restriction of the GPUs.
In comparison with~\cite{Tiddia2022}, the present work shows the performance improvements obtained using the new algorithm for network construction.
Additionally, a slight simulation performance gain can be noticed thanks to general optimizations to our state propagation implementation.

\begin{figure}[H]
    \centering
    \includegraphics[width=\linewidth]{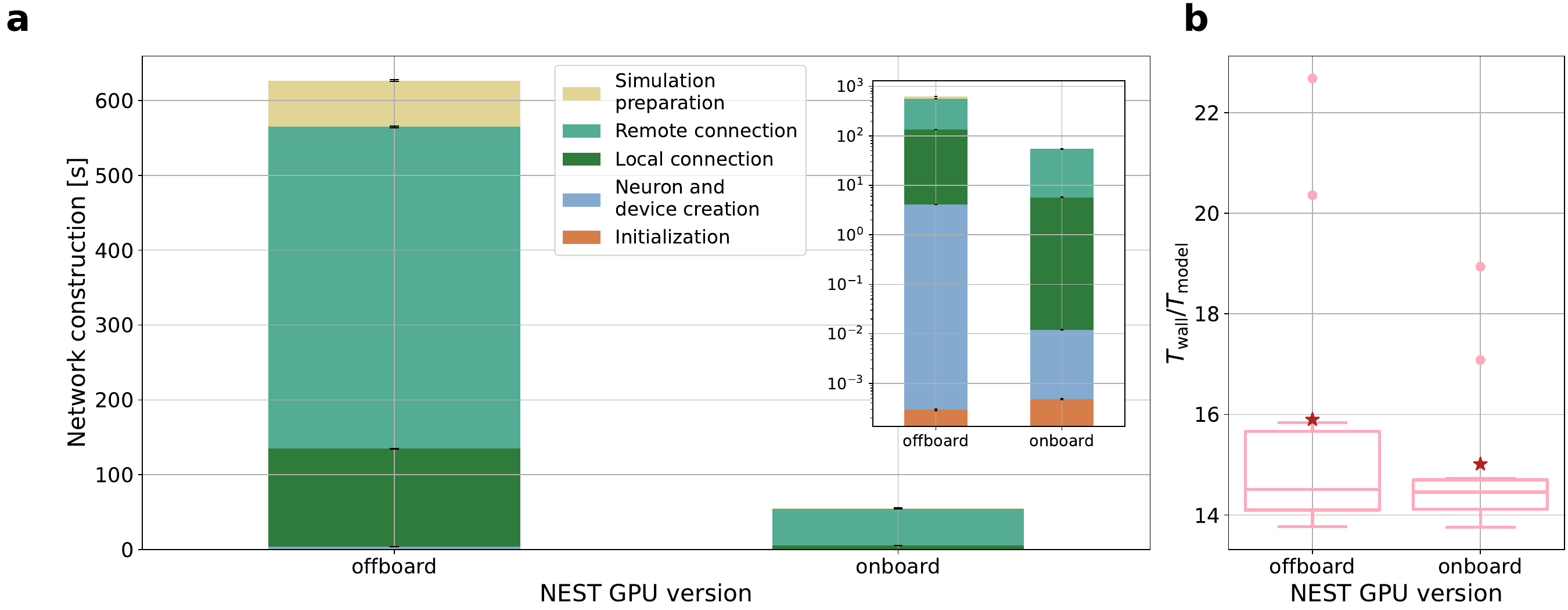}
    \caption{Comparison between performance of \textit{offboard} and \textit{onboard} versions on the simulation of the MAM in the metastable state. Simulations are performed on $32$ nodes on JUSUF (one NVIDIA V100 GPU per node). Mean data shown from averaging over $10$ simulations using different random seeds. Black error bars represent the standard deviations, which are barely visible due to the little variation across different simulations.
    (\textbf{a})\,Bar plot of the network construction time divided into its subtasks, which are shown in chronological order from bottom to top: 1) simulator initialization, 2) neuron and device creation, 3) local connection generation, 4) remote connection generation, and 5) the organization of data structures for spike delivery (\textit{simulation preparation}). Bar heights represent mean values, and black error bars standard deviations. At the top-right of the panel, an inset shows the phases of the network construction on a logarithmic scale to better visualize the contribution of initialization and neuron and device creation phases.
    (\textbf{b})\,Box plot of state propagation time measured as the real-time factor. The central line represents the median, whereas the box indicates the interquartile range (IQR). The whiskers extend up to $1.5\times$IQR, and data exceeding this value are represented as outliers. Red stars indicate the mean of the distributions.
    }
    \label{fig:MAM_simulations}
\end{figure}

Figure \ref{fig:MAM_simulations} compares \textit{offboard} and \textit{onboard} in the simulation of the MAM. Network construction time shown in panel \textbf{a} is more than $10$ times faster with the \textit{onboard} version.
In particular, network construction is $686.0 \pm 1.5$\,s for the \textit{offboard} and $55.5 \pm 0.1$\,s for the \textit{onboard} version. This time is divided into the subtasks of network construction. Connections creation, i.e., the most time-consuming phases of the network construction, shows $20\times$ and $9\times$ speedup for the \textit{onboard} version for local and remote connections creation, respectively. Neuron and device creation and simulation preparation, which have only a small contribution to the total network construction time also in the case of the \textit{offboard} version, are the phases benefiting from the largest speedup, respectively, of $350$ times and $50$ times. The state propagation time, represented in panel \textbf{b} with the real-time factor (see Equation \ref{eq:rtf}), is comparable between the two versions: $16.0 \pm 3.0$ and $15.0 \pm 1.7$ for \textit{offboard} and \textit{onboard} versions. The corresponding median values are $14.5$ and $14.4$, respectively. The differences in the extent of the two boxes and the outliers are related to the activity variability of the metastable state of the model, which may result in fluctuations in the spiking activity.

On the JUSUF cluster, each NVIDIA V100 GPU accommodates only a single MAM area. The NVIDIA A100 GPUs with $64$\,GB of GPU memory of the Leonardo Booster supercomputer~\cite{Turisini2024}, however, may host multiple areas on a single GPU card (see also Section \ref{sec:hardware}). 
An \textit{area packing} algorithm has been specifically designed for the NEST GPU implementation of the MAM model, which enables the simulation of the model using down to $8$ GPUs. These results are described in Appendix \ref{app:mam_area_packing}, and the algorithm is described in Section \ref{sec:area_packing}.

\subsection{Collective communication: scalable balanced network}
\label{sec:res_coll}
Weak scaling performance is assessed with a two-population random balanced network of excitatory and inhibitory neurons~\cite{Brunel2000}, similar to the one used in previous scaling studies~\cite{Albers2022}. Using random, fixed in-degree connectivity (multapses and autapses allowed), in principle, any neuron could be connected to any other one in that network. The average firing rate of excitatory and inhibitory populations is around $8$\,spikes per second.
Section \ref{sec:scalable_net} provides further details on that model.
The total network size is proportional to the product of a scale factor for the number of neurons per MPI process ($\texttt{scale}$ parameter) and the total number of MPI processes. The scale is constrained by the GPU memory available per MPI process.

We run simulations on the Leonardo Booster supercomputer~\cite{Turisini2024}, which is equipped with four NVIDIA A100 GPUs per cluster node, using one MPI process per GPU device, and employing collective MPI communication.
\textit{Collective protocols} allow the distribution of a communication payload across multiple processes. When network load across nodes is balanced and communication payloads are homogeneous, collective protocols provide latency, bandwidth, and overall communication speed benefits over point-to-point methods.
Results presented here correspond to $\texttt{scale}=20$ (i.e., $2.25 \times 10^5$ neurons and $2.53 \times 10^9$ synapses per GPU), which is the scale that would virtually enable the execution of such a model on the full Leonardo Booster. Table \ref{tab:model-size} shows the model size in terms of number of neurons and synapses as a function of the number of nodes and GPUs used in a simulation. Indeed, other choices are possible: Appendix~\ref{app:different_scales} reports node creation and connection time, as well as simulation preparation times, for the network model simulated at scales $10$ and $30$.

We perform simulations using from $32$ to $256$ nodes, i.e., up to $1,024$ GPUs. As CINECA regulations limit larger runs, we provide estimates for configurations requiring more nodes.
Since each MPI process carries out network construction and simulation preparation independently without the need for communication, it is possible to assess performance using a single compute node, with four MPI processes. For this, each process constructs its regular share of a large neuronal network in the absence of the remainder of the network.
While no state propagation occurs, this approach estimates GPU memory usage and construction time for large configurations at lower costs. We refer to such results as \textit{estimated} as opposed to \textit{simulated} outcomes.

For clarity, when evaluating simulated results, we analyze data from all processes active during simulation.
For instance, during simulations using $1,024$ processes, we analyze $1,024$ data samples of performance and memory usage metrics.
On the contrary, estimating the outcome  of $1024$ processes using only $4$ processes means we only analyze $4$ data samples.
In this case, each of the $4$ processes performs a \textit{dry-run} of network construction and simulation preparation as if $1,024$ processes were being used.
When estimating the results of stochastic connection rules, the performance and memory usage data reported by the limited number of processes performing the estimation are only a few samples of the actual distribution across the larger estimated number of processes.
Naturally, a single estimation run does not yield the full distribution of performance and memory usage. For statistical significance, we use $4$ processes per estimation run and average over $5$ estimation runs.
To obtain a complete estimate there exists literature on high-performance computing exploring statistical methods to predict memory and time requirements to run distributed applications~\cite{Reghenzani2020}; however, these methodologies are beyond the scope of the present work.

\begin{table}
\centering
\begin{tabular}{p{1.75cm} p{1.75cm} p{2.75cm} p{2.75cm}}
\toprule
\textbf{Nodes}&\textbf{GPUs}&\textbf{Neurons ($\times 10^6$)}&\textbf{Synapses ($\times  10^{12}$)}\\
\midrule
$32$  & $128$  & $28.8$  & $0.32$ \\
$64$  & $256$  & $57.6$  & $0.65$\\
$96$  & $384$  & $86.4$  & $0.97$\\
$128$ & $512$  & $115.2$ & $1.30$ \\
$192$ & $768$  & $172.8$ & $1.94$ \\
$256$ & $1,024$ & $230.4$ & $2.59$ \\
\bottomrule
\end{tabular}
\caption{\label{tab:model-size} Scalable balanced network model size (i.e., total number of neurons and synapses) as a function of the number of compute nodes. Each node is equipped with four GPUs, and the $\texttt{scale}$ of the model is set to $20$. The neuron number at \texttt{scale}$=1$ is $11,250$; the in-degree is fixed to $11,250$.}
\end{table}

In the following, we explore four alternative algorithms for balancing GPU memory usage and performance. Large-scale simulations occupy a relevant fraction of memory by data structures mapping remote source neurons to their local image neurons and outgoing connections on target MPI processes. To reduce GPU memory usage, parts of these structures can be stored in CPU memory, at the cost of slower network construction and simulation. The optimizations are ordered by increasing GPU memory usage, referred to as GPU memory levels, from level $0$ to level $3$. A detailed description of the different levels can be found in the Section \ref{sec:remote_conns}.

\begin{figure}[H]
    \centering
    \includegraphics[width=\linewidth]{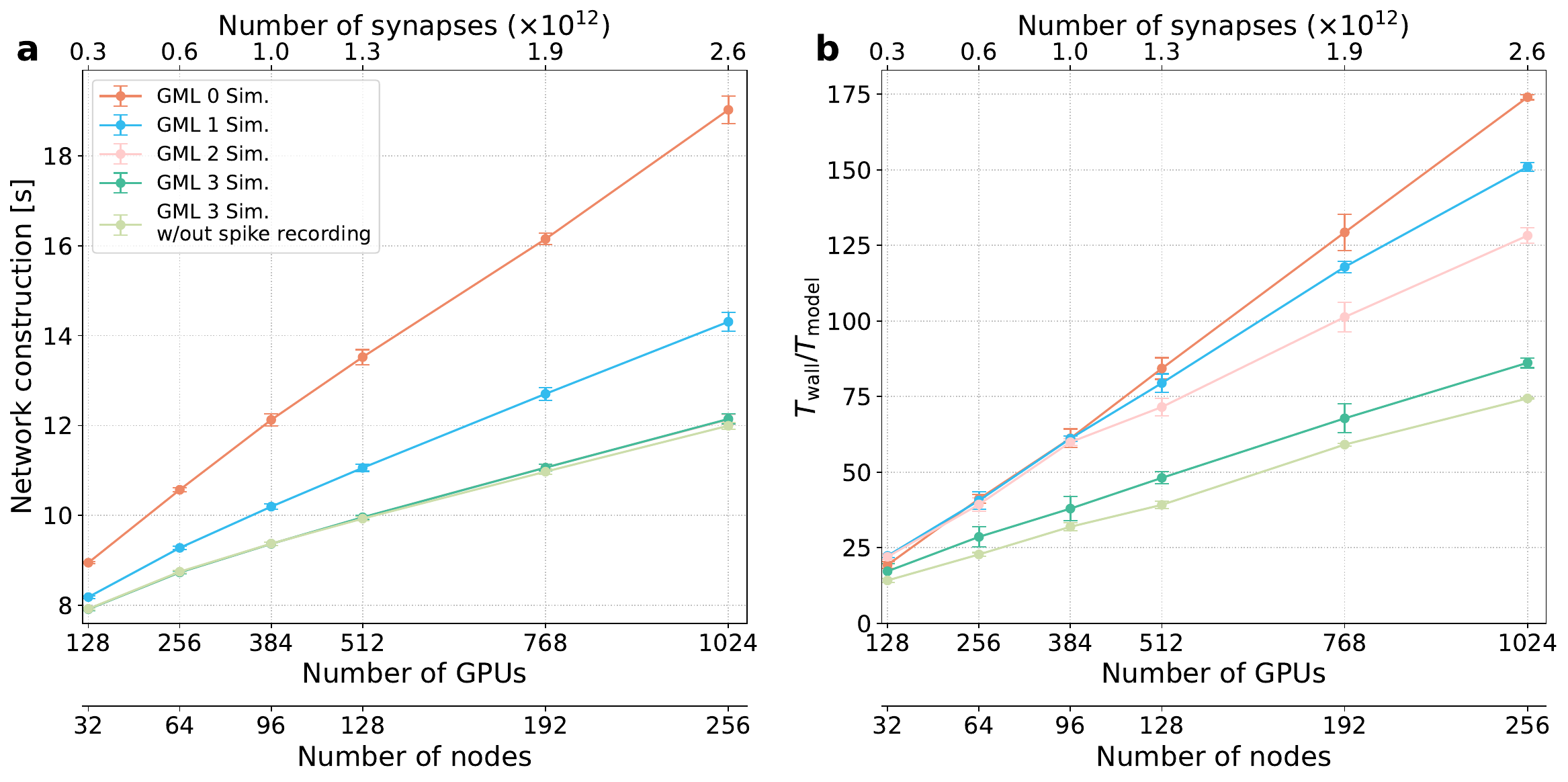}
    \caption{Network construction (\textbf{a}) and state propagation (\textbf{b}) times of the scalable balanced network model for the four GPU memory levels as a function of the number of cluster nodes (four GPUs per node). State propagation is measured as the real-time factor. The GPU memory level $3$ was also reported when spike recording was disabled. GPU memory levels $2$ and $3$ in the network construction overlap.}
    \label{fig:simulation_time}
\end{figure}

Figure~\ref{fig:simulation_time} summarizes network construction (panel~\textbf{a}) and state propagation (panel~\textbf{b}) time with increasing number of compute nodes for all GPU memory levels as a function of the number of cluster nodes.
As expected, higher GPU memory levels yield faster simulation speeds as data movement between CPU and GPU for communication is reduced due to the increasing availability of communication maps in GPU memory. 
For the GPU memory level $3$, which is the most efficient one in terms of time-to-solution, we also show the performance when spike recording is disabled, which show, on average a $20\%$ reduction in state propagation time.
Furthermore, network construction speed also benefits from higher GPU memory levels as within each GPU, creation and sorting of the communication maps is performed in parallel.
However, lower GPU memory levels are not without any gain, the lower the optimization, the lower the memory footprint of simulation-related data structures in GPU.

\begin{figure}[H]
    \centering
    \includegraphics[width=0.7\linewidth]{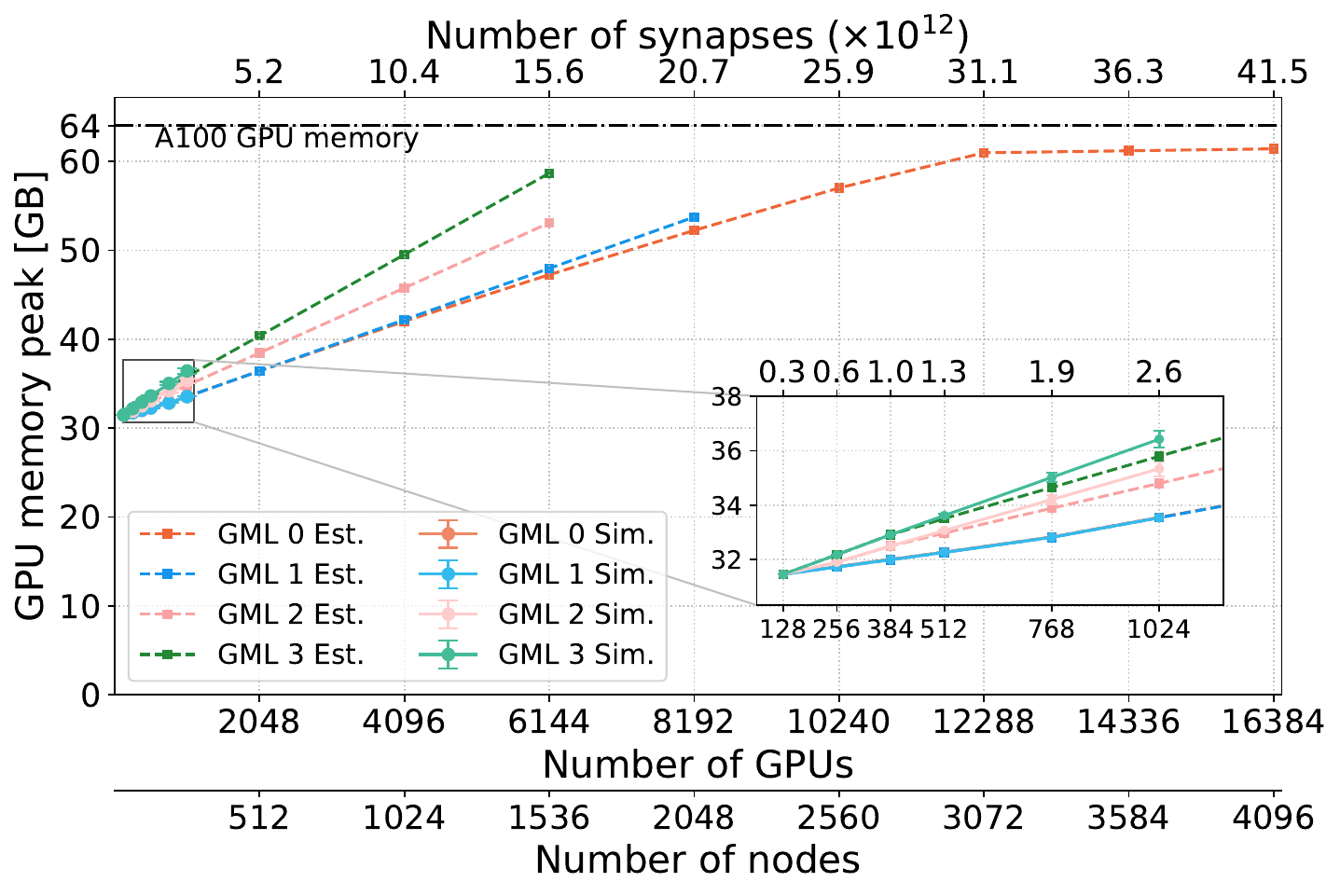}
    \caption{Peak memory usage per GPU for the scalable balanced network model simulation as a function of the number of cluster nodes for the four GPU memory levels. The upper horizontal axis shows the total number of synapses of the model, indicating the increasing network size.
    Square markers and dashed lines represent the estimation of the memory peak obtained using four MPI processes on a single node. Dot markers and continuous lines represent simulated configurations over five simulations using different random seeds. Error bars represent standard deviations.
    On the bottom right of the figure, an inset of the measured data is shown to facilitate comparison with the estimated values. Dash-dotted horizontal line represents the memory limit of one NVIDIA A100 GPU. For these configurations, the first two levels of GPU memory show compatible results, and the points overlap.}
    \label{fig:gpu_memory}
\end{figure}

Figure~\ref{fig:gpu_memory} shows the peak memory usage as a function of the number of compute nodes. Characterization of this peak is critical, as transient memory allocation can trigger out-of-memory errors and thus define the scalability limit of the simulation.
This illustrates the configurations that would succeed in constructing the network for each GPU memory level.

GPU memory level $0$ can reach a $4,096$-node configuration without exceeding the GPU memory of the A100 GPUs. From $3,072$ nodes onward, the GPU-memory peak plateaus due to the fixed number of in-degree per neuron of the network across all configurations.
Since the incoming connections are evenly distributed among MPI processes and GPU memory level $0$ copies to the GPU only the data of the remote neurons connected with the local ones, as the number of processes surpasses the in-degrees, the size of neuron maps copied does not increase anymore.
Other GPU memory levels require additional GPU memory, and the additional allocation for instantiating the network reaches the GPU memory limit.
In particular, the simulations with GPU memory levels 2 and 3 exhibit higher GPU peak memory usage than the respective estimates and also show variability across MPI processes.
This observation is explained by the non-deterministic allocation of temporary buffers needed for the network construction with the stochastic connectivity operations.

\begin{figure}[H]
    \centering
    \includegraphics[width=0.7\linewidth]{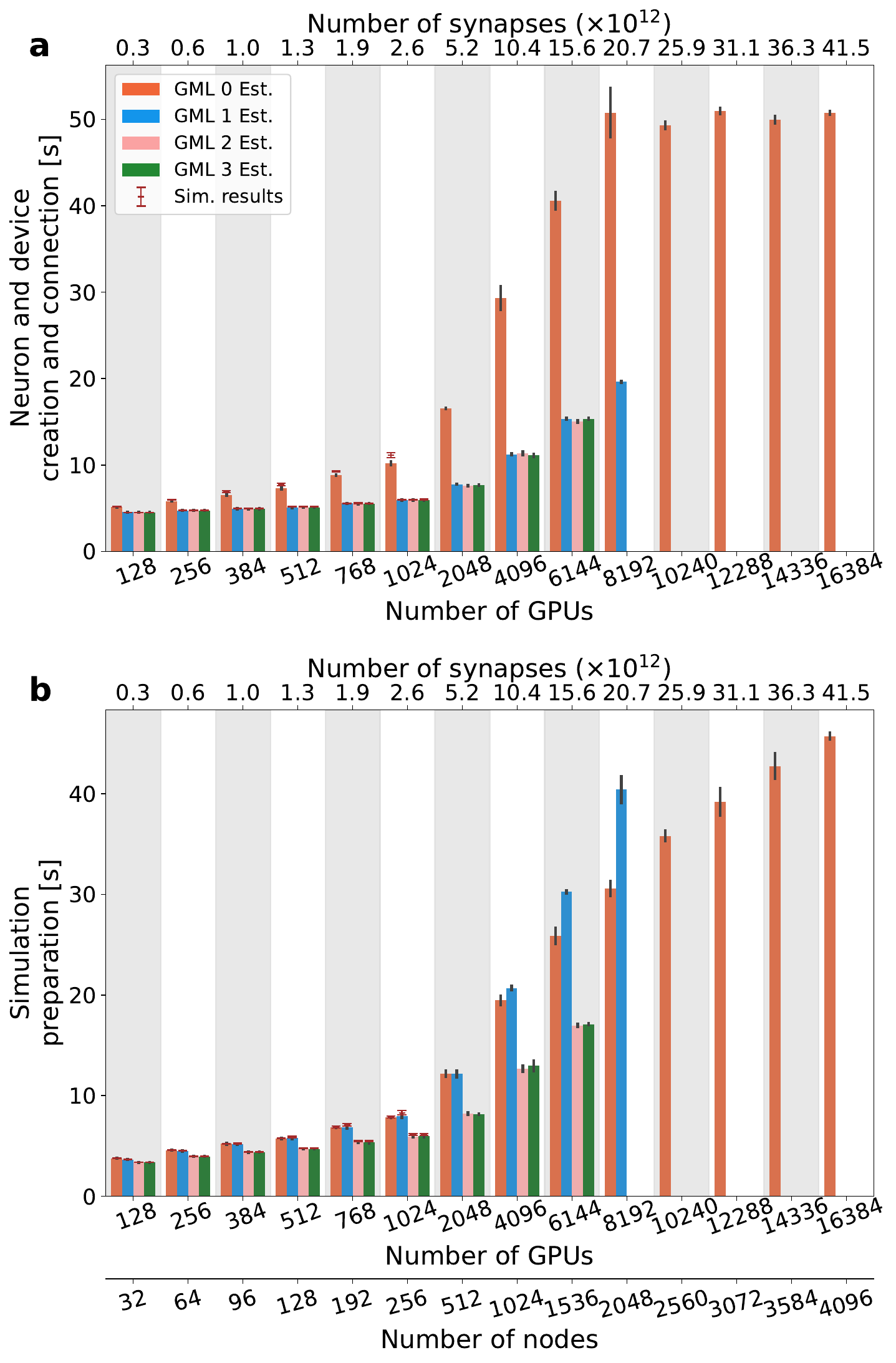}
    \caption{Network construction time divided into neuron creation and connection -- panel \textbf{a} contribution of the neuron and device creation and the local and remote connection generation subtasks -- and simulation preparation -- panel \textbf{b} organization of data structures for spike delivery subtask -- times of the scalable balanced network model evaluated across all MPI processes as a function of the number of cluster nodes for the three GPU memory levels. The bars represent an estimation using four MPI processes on a single node.
    The horizontal line markers show the time averaged across the MPI processes of five simulations with different seeds using the GPU memory levels. Errors for bars and horizontal line markers represent their standard deviation, with most of the values showing little variability.}
    \label{fig:-construct-and-preparation-time}
\end{figure}

Naturally, for each GPU memory level, we can also characterize the contribution of each network construction step to the overall time spent.
Figure~\ref{fig:-construct-and-preparation-time} splits network construction time into neuron and device creation, connection, and simulation preparation (organization of data structures for spike delivery) time as a function of the number of cluster nodes across all GPU memory levels. 
Here, initialization time is not considered due to its negligible contribution to the overall time: as shown in Figure \ref{fig:MAM_simulations}\textbf{a}, it is in the order of $10^{-2}$\,s and does not scale with network size.

For node and device creation and connection, in panel \ref{fig:-construct-and-preparation-time}\textbf{a}, GPU memory level $0$ has worse scaling compared to the higher GPU memory levels, which profit from the parallelization of the instantiation of the communication maps.

Moreover, a slight divergence is shown between simulated and estimated values for GPU memory level $0$, which increases with the number of nodes employed.
On the one hand, the probabilistic connection rule leads to a variability across processes, similar to its effect discussed with respect to the GPU peak memory usage for higher GPU memory levels.
On the other hand, GPU memory level $0$, using mostly the CPU for computing connections, is more susceptible to system jitter, which is commonly observed in large-scale systems~\cite{Reghenzani2020}.
Altogether, in such a large-scale environment, the cumulative effect of these sources of overhead on performance, not captured by the performance estimation, may become visible. We provided more details on the differences between simulated and estimated results in Appendix \ref{app:diff_est_sim}.

Panel \ref{fig:-construct-and-preparation-time}\textbf{b} shows a similar pattern for the simulation preparation step, with the exception of GPU memory level $1$, which now shows a similar scaling behavior to GPU memory level $0$.
This is expected, since, as reported in Section \ref{sec:opt_levels}, these memory levels have similar approaches, but while for level $0$ only the source neurons that have an image in the target MPI process are the ones that have a connection in the given MPI process, for level $1$ all source neurons have an image. With this, while globally GPU memory level $1$ is faster if the whole network construction time is considered (see Panel \ref{fig:simulation_time}a), the simulation preparation phase is slower than level $0$ since the operations to organize the connectivity become slower when the map increases in size.

Clearly, both instantiating and sorting communication maps have a deep impact during network construction and the advantage of parallelizing these operations in GPU is noticeable across scales.

\section*{Discussion}
The present study shows that GPU-based spiking neural network simulations greatly benefit from optimizing memory and speed during network construction. We devised and implemented a scalable network construction method directly in the GPU, without the need for MPI communication during this phase. The method can balance GPU memory consumption and performance, supporting both point-to-point and collective MPI communication, two modes whose efficiency depend on model structure and strongly influence state propagation time.
We report a more than ten times faster network construction of the large-scale multi-area model of the macaque vision-related cortex~\cite{Schmidt2018} compared to a previous implementation~\cite{Tiddia2022}.
Furthermore, we demonstrate weak scaling of a balanced random network model simulation up to $1,024$ GPUs, corresponding to a network of $230.4\times 10^6$ neurons and $2.59\times 10^{12}$ synapses, and provide a model to extrapolate performance beyond that.

At very-large model size, the GPU memory peak depends primarily on the number of connections allocated in each GPU, with other data structures, neuron parameters and spike recording having a negligible contribution. This is also true for neuron models with multiple receptors, already supported in NEST GPU, and would also be true for more complex neuron models, such as multicompartment neuron models, where one or more circular buffers are needed for each neuron compartment, similarly to what has already been done for the CPU version of NEST and the NEAT library~\cite{Wybo2021}.
Extrapolations suggest that the whole Leonardo Booster cluster (i.e., $3456$ nodes) could simulate a model with $3.5 \cdot 10^{13}$ connections --approximately $10\%$ of the connections in the human cortex--with an estimated node creation and connection time of around $50$\,s and a simulation preparation time of $42$\,s. 
However, such a network model is characterized by fully random connectivity. For a more realistic model of the cortex, more structured connectivity can further reduce simulation time if mapped efficiently onto the hardware. Indeed, several connections are lateral, with a decay length in the order of a few hundred micrometers~\cite{Boucsein2011}. Moreover, only a limited fraction of neurons project to remote areas~\cite{Rosen2022}. From a simulation perspective, lateral and long-range connections could be handled differently: a hybrid scheme combining point-to-point communication for local spikes and collective MPI communication for long-range projections would be particularly advantageous.
Going from the load balancing round-robin distribution of neurons to network-structure-aware model representations on the hardware requires advanced mapping strategies. In our current implementation, load balancing needs to be handled by the user, which is sub-optimal given the complexity of the physical hardware networking topology unique to each compute cluster, and could be addressed in the future. Moreover, the \textit{area packing} algorithm developed here for the MAM implementation could be generalized to further models with modular structure. Most of such models are driven by anatomical and physiological data that are increasingly available and improving in resolution. In particular, human brain atlases and parcellations spanning tens to hundreds, up to around a thousand parcels, are commonly used together with inter-parcel connectivity matrices and with inter-parcel distance information (e.g., tract length or signal/conduction delay). These features can be leveraged by partitioning algorithms to optimize the distribution of the workload across processes~\cite{Hagmann2008,Lober2026}.
Furthermore, there exist spiking neural network simulator implementations that employ graph partitioning algorithms to balance the compute load for distributed simulations~\cite{Wang2024, Qu2023}.
These could serve as algorithmic references and points of comparison for future load balancing implementations in NEST GPU, however such studies would require accompanying validation of spiking statistics across simulations to verify that network dynamics match.

Further challenges related to network connectivity include, for example, dynamic connectivity changes and loading of existing connectomes: first, modeling structural plasticity requires the efficient creation and removal of synapses during the simulation at runtime. A flexible framework for structural plasticity on a single GPU has recently been published~\cite{Knight2026}. Algorithms for distributed simulations across multiple GPUs are subject to future work.
Second, large-scale models with data-driven connectivity (e.g., \cite{Gandolfi2023}) would benefit from efficient routines for loading complete connectomes from file instead of the current practice of initiating the network graph every time prior to the simulation phase. Future work introducing support for the SONATA format~\cite{Dai2020}, together with strategies for storing and subsequent reloading of network connectivity could enhance network construction capabilities accordingly.

Aside from those challenges, this work shows the capability for NEST GPU to meet large-scale simulation requirements. Memory footprints shown in Figure~\ref{fig:gpu_memory} suggest that our reference implementation can scale to the largest current HPC systems. Emerging exascale machines are composed of fewer than ten thousand nodes, with each node providing high computational density. It could then host full machine runs and increase the neural network density at each node. The exascale supercomputer JUPITER Booster \cite{Herten24_1} uses NVIDIA GH200 super-chips (\url{https://resources.nvidia.com/en-us-grace-cpu/grace-hopper-superchip}, last visited: 05.09.2025). With this hardware reference, simulations of the scalable balanced network model would become possible at higher scale and almost double the number of compute nodes with respect to the Leonardo Booster, leading to simulations of networks with the same order of magnitude of neurons and synapses of the human cortex, while allocating individual synapse information.

\section*{Methods}\label{sec:methods}
\subsection{Remote connections}\label{sec:remote_conns}
The MPI protocol allows for the management of simulations running on multi-GPU systems and on GPU clusters in a conceptually similar way. In both cases, the communication of spikes and other types of data between GPU devices is handled via MPI, regardless of whether the devices belong to the same node or to different ones in the cluster. Here, we use the term {\it multi-GPU simulation} to refer to any simulation involving more than one GPU, regardless of their actual physical location.

In such simulations, it is useful to distinguish between {\it local} connections, in which both the source and target neurons are located in the same MPI process, and {\it remote} connections, which span different MPI processes. 
Local connections are created by the invocation of the \texttt{Connect}
method~\cite{Golosio23_9598}, while remote connections are generated through a dedicated \texttt{RemoteConnect} method, which will be described in this section.
The strategy used in this work for managing remote connections involves representing the source neuron in the target MPI process using a virtual {\it image neuron} (or {\it proxy}).
Each local neuron can be uniquely indexed within an MPI rank $\sigma$, for the following definitions, the set $\mathbf{N}_\sigma = \{0, ..., M_\sigma - 1\}$, where $M_\sigma$ is the number of neurons in the rank, $\mathbf{N}$ is the collection of these indexes, $\mathbf{N}_i$ is a neuron index, and the cardinality $|\mathbf{N}_\sigma| = M_\sigma$.
Image neurons are not explicitly instantiated in memory but are identified by separate local indexes in the target MPI process.

This representation allows for remote connections to be built in the target process using the same mechanisms as local connections, with the local index of the image neuron serving as the source. Efficient management of remote connection creation and spike communication relies on mapping structures that associate the index of the remote source neuron in the source MPI process to the corresponding local index of the image neuron in the target MPI process, where the connection is allocated.

The details of how those maps are created and used are slightly different depending on which MPI communication approach is used, i.e., {\it point-to-point} or {\it collective}. 
In the following paragraphs, we will describe them, analyze their differences and discuss their advantages and disadvantages.

\subsubsection{Point-to-point MPI communication}
This approach is particularly useful when each source neuron of a population allocated in a certain source MPI process has a large number of connections directed to the same target MPI process. For example, this could be the case in models where the distribution of neurons among MPI processes takes spatial location into account, for connections between MPI processes representing adjacent brain regions.

Maps associate the source neurons of remote connections with their images in the target MPI process, defined as follows:
\begin{itemize}
    \item In the target MPI process, there is one map for each possible source MPI process. Those maps are pairs of the type $(R_{\tau, \sigma, i}, L_{\tau, \sigma, i})$
    where $R_{\tau, \sigma, i}$
    is the remote source neuron index, i.e., the index of the source
    neuron in the MPI process where it is allocated, and
    $L_{\tau, \sigma, i}$
    is the local image neuron index, i.e., the index of the image neuron in the target MPI process where the connection is stored. $\tau$ is the target MPI process rank, $\sigma$ is the source MPI process rank,
    and $i \in \{0, ..., N_{\tau, \sigma} - 1\}$
    is the index that represents the position in the map 
    ($N_{\tau, \sigma}$ being the number of elements in this map).
    The pairs are stored on each target MPI process using two arrays per source MPI process.
    These arrays are organized in fixed-size blocks that are allocated dynamically in order to use GPU memory efficiently. Note that the arrays of index $\tau$ are allocated only in the target MPI process of rank $\tau$.
    
    \item In the source MPI process, there is one sequence for each possible target MPI process,
    $S_{\tau, \sigma, i}$, where $\tau$ , $\sigma$ and $i$ follow the previous definitions and $S_{\tau, \sigma, i}$ is the local source neuron index.
\end{itemize}

Clearly,
\begin{equation}\label{eq:mapalign}
    S_{\tau, \sigma, i} = R_{\tau, \sigma, i} ;
\end{equation}
however, the first array is stored in the source MPI process
(rank $\sigma$), while the second is stored in the target MPI process (rank $\tau$).
In the following text, we will use boldface characters to represent sequences such as:
\begin{equation}
\mathbf{R}_{\tau, \sigma} = 
( R_{\tau, \sigma, 0}, ... ,
R_{\tau, \sigma, N_{\tau, \sigma} - 1} )
\end{equation}
with similar definitions for
$\mathbf{L}_{\tau, \sigma}$ and
$\mathbf{S}_{\tau, \sigma}$.

Map elements are sorted in ascending order with respect to the source neuron index, i.e.,
\begin{equation}\label{eq:sortedmaps}
S_{\tau, \sigma, i} \geq
S_{\tau, \sigma, i-1}
\quad \forall i \in \{1, ..., N_{\tau, \sigma} - 1\}
\end{equation}
Additionally, for each neuron 
allocated in the MPI process $\sigma$, i.e.
for each index
$s = 0, ..., M_\sigma -1$,
where $M_\sigma$ is the number of neurons in $\sigma$,
we create the sequence
$T_{\sigma, s, j}$,
of the MPI processes in which $s$ has an image, where
$j = 0, ..., J_{\sigma, s} - 1$
is the index of the elements in the sequence and $J_{\sigma, s}$
is their number, i.e. the number of MPI processes on which $s$ has an image.
This sequence is combined with the positions 
$P_{\sigma, s, j}$
of the pairs (source-neuron index, image-neuron index) in the
$(R, L )$ maps.
Specifically, each element $P_{\sigma, s, j}$ stores the index corresponding to the position of the pair within the $(R, L )$ map in the target MPI process.
The MPI processes $T_{\sigma, s, j}$
are the ones that need to be properly informed when the node $s$ emits a spike.
Figure \ref{fig:p2p_routing} illustrates how the sequences 
$\mathbf{T}_{\sigma, s}$,
$\mathbf{P}_{\sigma, s}$
and
$\mathbf{L}_{\tau, \sigma}$
are used to communicate the spikes and to transmit them across the remote connections.
Appendix \ref{app:p2p} shows how the maps $(R_{\tau, \sigma, i}, L_{\tau, \sigma, i})$
and the sequences $\mathbf{L}_{\tau, \sigma}$, $\mathbf{T}_{\sigma, s}$ and $\mathbf{P}_{\sigma, s}$ are organized in GPU memory, and describes in a more detailed example how those structures are used to communicate remote spikes.

It should be noted that in the source MPI process we store the position of the image neuron in the map, rather than its index in the target MPI process. In fact, as explained below, this position can be determined during network construction without communication between processes, and this allows to considerably reduce the network creation time.
The sequences
$\mathbf{S}_{\tau, \sigma}$,
stored in the source MPI process
$\sigma$, must be aligned with the sequences 
$\mathbf{R}_{\tau, \sigma}$ and
$\mathbf{L}_{\tau, \sigma}$
stored in the target MPI process $\tau$ (Eq. \ref{eq:mapalign}).
A difficulty in keeping this alignment arises from the fact that the maps are updated dynamically every time a remote connect function is instantiated, and in many connection rules, the source and target neurons of the connections are determined from probabilistic distributions using pseudo-random numbers.
A possible solution could be to create the maps only in the target MPI processes, and at the end of the network construction phase, just before the simulation of the network dynamics, to communicate them to the source MPI processes. However, for large networks, this communication has a deep impact on the network construction time \cite{Tiddia2022}.
The approach used in this work relies on a source process variant of the \texttt{RemoteConnect}
method that is instantiated in the source MPI process, while the real method is used in the target process to create the connections.
The method variant's sole function is to generate in the source MPI process the same source neuron indexes of the new connections generated in the target MPI process, based on the specified connection rule.
These indexes are used to update the
$\mathbf{S}_{\sigma, \tau}$
sequence in the source process and the
$\mathbf{R}_{\tau, \sigma}$ and
$\mathbf{L}_{\tau, \sigma}$
sequences
on the target process each time the \texttt{RemoteConnect} method is instantiated.
Generating the same source neuron indexes allows those sequences to be aligned (Eq.~\ref{eq:mapalign}) without the need for MPI communication during the network construction phase.
Since some connection rules~\cite{Senk2022} require that the indexes of the source neurons be extracted using pseudo-random numbers according to given probabilistic rules, an array of generators,
\texttt{RNG}$_{\sigma, \tau}$, is used to generate these sequences. The
\texttt{RNG}$_{\sigma, \tau}$ generator starts from the same seed in the source MPI process $\sigma$ and in the target MPI process $\tau$, and is used exclusively to generate the indexes of the source neurons of the remote connections, so that these sequences are always aligned between the two processes.

\subsubsection{Collective MPI communication}
The method used to distribute spikes in groups of MPI processes with collective calls is similar to that used for point-to-point communication, however, an additional structure is used in this case: for each process in a given group, an array containing the indexes of source neurons with outgoing connections to target neurons of other processes in the same group, $H_{\alpha, \sigma, i}$, is created in each process of the same group.
Here
$\alpha$ is the group index,
$\sigma$ is the rank of the process in which the source neuron is located,
$i \in \{0, ..., N_{\alpha, \sigma}-1\}$
is the position in the array,
$N_{\alpha, \sigma}$
being the array size,
and $H_{\alpha, \sigma, i}$ is the index of the source neuron in process
$\sigma$.
This array is sorted in ascending order, i.e.,
$H_{\alpha, \sigma, i} \geq
H_{\alpha, \sigma, i-1}$
$\forall i \in \{1, ..., N_{\alpha, \sigma}-1\}$, and allows to efficiently keep track of remote source neurons already used in previously created connections. A second array, $I_{\alpha, \tau, \sigma, i}$, aligned with the previous one, represents the local index, in the process of rank $\tau$ in the MPI group $\alpha$, of the image neuron corresponding to the remote source neuron $H_{\alpha, \sigma, i}$, if the latter node has outgoing remote connections directed to target neurons allocated in the process $\tau$, while by convention it is set equal to -1 if the source neuron does not have an image neuron in $\tau$. Furthermore, while for point-to-point communication each source neuron was associated with a sequence of MPI target processes, in this case each node $s$ is associated to the sequence 
of MPI groups
$G_{\sigma, s, j}$
in which it has an image
and, for each group, to the corresponding position
$i \gets Q_{\sigma, s, j}$
of the node in the sequence
$H_{\alpha, \sigma, i}$.
\\
$G_{\sigma, s, j}$ are the groups to which the source neuron must report that it has produced a spike when it emits one.
The collective communication function used in this work is \texttt{MPI\_Allgather}.
With this function, each process in an MPI group communicates the spikes of its source neurons simultaneously to all other processes in the same group, while listening for those sent by others.
Figure~\ref{fig:collective_routing} illustrates how the spikes are communicated within MPI groups through collective communication.

\subsubsection{Creation of the remote connection structures for point-to-point MPI communication}
Remote connections can be created dynamically by a method of the simulator class,
$\texttt{RemoteConnect}
(\sigma, \mathbf{s},
\tau, \mathbf{t}, p,
\mathcal{C}, \mathcal{D}, \alpha)$
where
\begin{itemize}
\item $\sigma$ is
 the source MPI process rank;
 \item 
 $\mathbf{s} = ( s_0, ..., s_{N_\mathbf{source} - 1} )$
is the array of the indexes of the source neurons, $N_\mathbf{source}$ is the number of source neurons used in the call;
\item $\tau$ is
 the target MPI process rank;
 \item 
 $\mathbf{t} = ( t_0, ..., t_{N_\mathbf{target} - 1} )$
is the array of the indexes of the target neurons, $N_\mathbf{target}$ is the number of target neurons used in the call;
\item $p$ is the receptor port index;
\item $\mathcal{C}$ is the connection rule dictionary \cite{Senk2022};
\item $\mathcal{D}$ is the synaptic parameters dictionary;
\item $\alpha$ is the MPI group;
a special value
($\alpha = -1$ in our work)
for this argument is used to create remote connections for point-to-point MPI communication.
\end{itemize}
Special cases arise when
$\mathbf{s}$ and/or
$\mathbf{t}$
are sequences of consecutive integers, i.e.,
$s_i = s_0 + i$
and/or
$t_i = t_0 + i$.
In particular, the case where
$\mathbf{s}$
is such a sequence allows the
sequences
$\mathbf{S}_{\tau, \sigma}$,
$\mathbf{R}_{\tau, \sigma}$ and
$\mathbf{L}_{\tau, \sigma}$
to be updated more quickly, thanks to the fact that these sequences are sorted according to the index of the source neuron (Eqs. \ref{eq:mapalign} and \ref{eq:sortedmaps}).

The \texttt{RemoteConnect} method launches the CUDA kernels~\cite{cuda}
that build  the structures that are used to extract the spikes that must be communicated remotely from the GPU memory of the source process and create the connections~\cite{Golosio23_9598} and the structures  that are used to deliver  the spikes in the GPU memory of the target MPI process.
As we will see in Section \ref{sec:opt_levels}, in order to save GPU memory some of these structures can optionally be stored in CPU memory, at the expense of increasing network creation time and dynamics simulation time.
To account for this possibility, GPUDirect communication was not implemented in this work, although on some hardware it could significantly reduce communication time between GPUs. Interprocess communication was based on the MPI C/C++ libraries.

To represent the local indexes of image neurons corresponding to the remote neurons $\mathbf{s}$,
the \texttt{RemoteConnect} method initially allocates GPU memory to hold a temporary array,
$\mathbf{l} = ( l_0, ..., l_{N_\mathbf{source} - 1} )$
with the same size as
$\mathbf{s}$.

Many connection rules prescribe extraction of the source and/or target neuron index of each connection from the sequences
$\mathbf{s}$ and $\mathbf{t}$
in a nondeterministic manner through pseudo-random number generators (e.g., \textit{random, fixed in-degree} and \textit{random, fixed total number}) \cite{Senk2022}.
In some cases, the source neurons that are used in the newly-created connections are only a fraction of the set specified by the \texttt{RemoteConnect} call.
In order to save GPU memory, it can be convenient to flag the nodes that are actually used, i.e., nodes that have at least one outgoing connection in the target MPI process; only these nodes should be inserted in the maps that associate remote source neurons to local images.
This step is not necessary for connection rules that always use all source neurons (as \textit{all-to-all}, \textit{one-to-one}, \textit{random, fixed out-degree}). Furthermore, the increase in network construction time associated with this step may not be justified if the fraction of source neurons that are not used is small.
The decision on whether to use this step is a compromise between GPU memory usage and building time; for connection rules in which the source neuron indexes are extracted randomly (e.g. \textit{random, fixed in-degree} and \textit{random, fixed total number}) this decision can be based on the heuristic approach of using a threshold $\xi$ on the ratio between the estimated number of newly-created connections and the size of the set of source neurons specified by the
\texttt{RemoteConnect} call
(e.g., $K_\text{in} \times N_\mathbf{target} / N_\mathbf{source}
< \xi$
for the \textit{random, fixed in-degree} connection rule).
In this work we used a threshold $\xi = 1$, unless differently specified. 
In case the flagging approach is used, on both the source and target MPI processes, a temporary array of boolean values is created in GPU memory, $b_i$, with 
$i \in \{0, ..., N_\mathbf{source} -1\}$ and with all elements initially set to
$\texttt{false}$.

The connections outgoing from the image neurons are created in the target MPI process using a 
\texttt{Connect} method,
in similar way as for local connections~\cite{Golosio23_9598}, however instead of using
the actual set of source neuron indexes
specified by the \texttt{RemoteConnect} call,
the integers from $0$ to $N_\mathbf{source} -1$
 are temporarily used.
This is advantageous because in a later step it will be necessary to set the source neuron of each new connection to the index of the corresponding local image neuron, and as we will see later this index can be quickly obtained by knowing its position in the $\mathbf{s}$ sequence.
Furthermore, in case the source neuron indexes have to be extracted using a probabilistic rule, the
aligned random number generator array
\texttt{RNG}$_{\sigma, \tau}$
is used for this purpose.
As mentioned previously, the source MPI process uses a variant of the \texttt{RemoteConnect} method, which
performs only the extraction
of the source neuron indexes.
 For each new connection, we set
 $b_i = \text{true}$, with the index $i \in \{0, ..., N_\mathbf{source} -1\}$
 that identifies the source neuron of the connection in the sequence
 $\mathbf{s}$.
In this way, only the source neurons that are actually used by the new connections are labelled as
$\texttt{true}$.
An array $\tilde{\mathbf{u}}$ is then created containing the indexes of the source neurons $s_i$
for which $b_i = \texttt{true}$:
$\tilde{\mathbf{u}} = 
(i : b_i = \texttt{true})$,
and a subarray,
$\tilde{\mathbf{s}}$,
is then extracted from $\mathbf{s}$ using these indexes:
\begin{equation}
\tilde{s}_j \coloneqq s_{\tilde{u}_j}
\end{equation}
with $j \in \{0, ..., 
\widetilde{N}_\mathbf{source}\}$,
where $\widetilde{N}_\mathbf{source}$ is the number of source neurons actually used in the new connections.
$\tilde{\mathbf{u}}$
and $\tilde{\mathbf{s}}$
are sorted in ascending order with respect to the values of the elements of $\tilde{\mathbf{s}}$.

Each element of the array
$\tilde{\mathbf{s}}$
is searched in $R_{\tau, \sigma, i}$
to see if the source neuron it represents is already mapped to
an image neuron in the target MPI process.
In this case, the index of the image neuron obtained from the map
is assigned to the corresponding element of the array
$\mathbf{l}$. In formulas:
\begin{equation}
\text{If } \exists i \text{ such that } R_{\tau, \sigma, i} = \tilde{s}_j,
\quad \text{then set } l_{\tilde{u}_j} \coloneqq L_{\tau, \sigma, i} .
\end{equation}
Otherwise, if no entry for $\tilde{s}_j$ exists in the map, 
a new (virtual) image neuron is created in the target MPI process $\tau$.
This is done by appending the remote source neuron index $\tilde{s}_j$ to the 
sequence
$\mathbf{R}_{\tau,\sigma}$, appending the current number of nodes $M_\tau$ 
to the sequence
$\mathbf{L}_{\tau,\sigma}$, and then increasing $M_\tau$ by one:
\begin{equation}
\begin{aligned}
\mathbf{R}_{\tau,\sigma} &\gets (\mathbf{R}_{\tau,\sigma}, \tilde{s}_j), \\
\mathbf{L}_{\tau,\sigma} &\gets (\mathbf{L}_{\tau,\sigma}, M_\tau), \\
l_{\tilde{u}_j} &\coloneqq M_\tau, \\
M_\tau &\gets M_\tau + 1 .
\end{aligned}
\end{equation}
After completing this procedure, the map
$(R_{\tau,\sigma, i},
L_{\tau,\sigma, i})$
is again sorted in ascending order with respect to the values of
$\mathbf{R}_{\tau,\sigma}$ (Eq. \ref{eq:sortedmaps}).

Finally, the source neuron indexes of all new connections, for which the positions in the $\mathbf{s}$ sequence  were temporarily used, are replaced with the corresponding image neuron indexes, $l_i$.

In the source MPI process, the source process variant of the
\texttt{RemoteConnect} method
only generates the sequence of source
node indexes of the new connections,
using the pseudo-random number 
generator
\texttt{RNG}$_{\sigma, \tau}$,
aligned with the target process,
as discussed previously.
Those indexes are used to
build the array 
$\tilde{\mathbf{s}}$
with the same procedure
described for the target MPI process.
The elements of this array are used to
update the sequence
$\mathbf{S}_{\tau, \sigma}$:
if no entry 
for $\tilde{s}_j$ exists in this sequence, then this value is appended to it:
\begin{equation}
\mathbf{S}_{\tau, \sigma} \gets (\mathbf{S}_{\tau, \sigma}, \tilde{s}_j),
\end{equation}
After this procedure is completed,
$\mathbf{S}_{\tau, \sigma}$ is sorted in ascending order.
This ensures that
$\mathbf{S}_{\tau, \sigma}$ and
$\mathbf{R}_{\tau,\sigma}$
are always aligned (Eq. \ref{eq:mapalign}).

In the simulation preparation phase,
the sequences $\mathbf{S}_{\tau, \sigma}$
are used to extract,
for each source neuron index $s$,
a sequence $\mathbf{T}_{\sigma, s}$ of the target MPI processes
where $s$ has an image neuron,
and to associate to each element
of this sequence,
$T_{\sigma, s, j}$,
the corresponding
position $P_{\sigma, s, j}$
of the source neuron index
in the sequence
$S_{\tau, \sigma, i}$.
More specifically,
for each target process rank, $\tau$,
and for each index $i$
in the range $[0, N_{\tau, \sigma}-1]$,
we call $s$ the index of the source neuron represented by the $i$-th element of the sequence $\mathbf{S}_{\tau, \sigma}$
\begin{equation}
s \gets S_{\tau, \sigma, i}
\end{equation}
and we append the target process rank,
$\tau$, and the index $i$
to the sequences
$\mathbf{T}_{\sigma, s}$ and 
$\mathbf{P}_{\sigma, s}$,
respectively:
\begin{equation}
\begin{aligned}
    \mathbf{T}_{\sigma, s}
    &\gets 
    (\mathbf{T}_{\sigma, s}, \tau) \\
    \mathbf{P}_{\sigma, s}
    &\gets
    (\mathbf{P}_{\sigma, s}, i) \\
\end{aligned}
\end{equation}
The position $i$ of the source neuron will be the same in the map
$(R_{\tau,\sigma, i},
L_{\tau,\sigma, i})$ 
in the target MPI process, because these sequences are aligned (Eq. \ref{eq:mapalign}).
As discussed previously, to deliver a spike from the source neuron $s$
to the target process
$\tau \gets T_{\sigma, s, j}$,
its corresponding positions 
$i \gets P_{\sigma, s, j}$
is sent to this process,
where it is used  to extract the corresponding image neuron index
$L_{\tau,\sigma, i}$,
and thus to deliver the spike
through the connections outgoing from this node.

\subsubsection{Creation of the remote connection structures for collective MPI communication}
Maps that associate remote source neurons to their local images
in the target MPI process are also created for collective MPI spike communications, except that in this case an additional index,
$\alpha$, is needed to represent the MPI group:
\begin{equation}
( R_{\alpha + 1, \tau, \sigma, i},
L_{\alpha + 1, \tau,\sigma, i} )
\end{equation}
is the map that associates the indexes of the source neurons of the MPI process
of rank $\sigma$ in the $\alpha$ MPI group to the corresponding indexes of the image neurons in the target MPI process of rank $\tau$ in the same group.
As we wrote previously, by convention,
the value $\alpha = -1$ is used to indicate point-to-point communication.
Thanks to this convention, the definitions of $R$ and $L$ given for point-to-point communications can be merged with those used for collective communications by setting
\begin{equation}
\begin{aligned}
R_{0, \tau, \sigma, i}
 &\coloneq R_{\tau, \sigma, i} \\
L_{0, \tau,\sigma, i}
&\coloneq
L_{\tau,\sigma, i} \\
\end{aligned}
\end{equation}
The procedure used to construct these maps and to create image neurons and remote connections is the same as described for the point-to-point case.

However, for collective communication, it is not necessary to create the sequence
$\mathbf{S}_{\tau, \sigma}$ in the
source MPI process.

Instead, the set $\mathcal{H}_{\alpha, \sigma}$ of the source neurons of the process of rank $\sigma$ in the group
$\alpha$ passed as arguments to the \texttt{RemoteConnect} calls
is created in all MPI processes of this group:
for each invocation of the
\texttt{RemoteConnect} method,
$\mathcal{H}_{\alpha, \sigma}$ is updated according to the rule:
\begin{equation}
\mathcal{H}_{\alpha, \sigma} \gets
\mathcal{H}_{\alpha, \sigma}
\cup
\{s_0, ...,
s_{N_\mathbf{source} - 1}\}
\end{equation}
In the simulation preparation phase, the elements of $\mathcal{H}_{\alpha, \sigma}$
are placed in a sequence,
$\mathbf{H}_{\alpha, \sigma}$,
sorted in ascending order.
\begin{equation}
\mathbf{H}_{\alpha, \sigma} = \mathbf{sort}
(\mathcal{H}_{\alpha, \sigma})
\end{equation}
In each process $\tau$ in the MPI group $\alpha$, an array,
$\mathbf{I}_{\alpha, \sigma, \tau}$, is then created with the same dimensions as
$\mathbf{H}_{\alpha, \sigma}$.
This array is intended to contain the local image neuron indexes corresponding to the source neurons inserted in
$\mathcal{H}_{\alpha, \sigma}$
that have an image in the process
$\tau$.
The elements of
$\mathbf{I}_{\alpha, \sigma, \tau}$
are initialized to -1; this value remains unchanged for nodes that do not have an image.
The
$(R_{\alpha+1, \tau, \sigma, i}, L_{\alpha+1, \tau, \sigma, i})$
map is then used to assign the values
of $\mathbf{I}_{\alpha, \sigma, \tau}$:
\begin{equation}
\text{If } \exists i \text{ such that } R_{\alpha+1, \tau, \sigma, i} = 
H_{\alpha, \sigma, j},
\quad \text{then set }
I_{\alpha, \sigma, \tau, j}
 \coloneqq L_{\alpha+1, \tau, \sigma, i} .
\end{equation}
The pairs
$(H_{\alpha, \sigma, i} , 
I_{\alpha, \sigma, \tau, i})$
can be used as a map
that associates
the remote source neuron index to
its local image neuron index.

In the source MPI process $\sigma$,
the sequences
$\mathbf{H}_{\alpha, \sigma}$
are used to extract,
for each source neuron index $s$,
a sequence $\mathbf{G}_{\sigma, s}$ of the MPI groups
where $s$ has an image neuron,
and to associate to each element
of this sequence,
$G_{\sigma, s, j}$,
the corresponding
position 
$Q_{\sigma, s, j}$
of the source neuron index
in the sequence
$H_{\alpha, \sigma, i}$, through the following procedure:
for each MPI group, $\alpha$,
and for each index $i$
in the range $[0,
\mathbf{size}
(\mathbf{H}_{\alpha, \sigma}) - 1]$,
we call $s$ the index of the source neuron represented by the $i$-th element of the sequence $\mathbf{H}_{\alpha, \sigma}$
\begin{equation}
s \gets H_{\alpha, \sigma, i}
\end{equation}
and we append the MPI group index
$\alpha$ and the index $i$
to the sequences
$\mathbf{G}_{\sigma, s}$ and 
$\mathbf{Q}_{\sigma, s}$,
respectively:
\begin{equation}
\begin{aligned}
    \mathbf{G}_{\sigma, s}
    &\gets 
    (\mathbf{G}_{\sigma, s}, \alpha) \\
    \mathbf{Q}_{\sigma, s}
    &\gets
    (\mathbf{Q}_{\sigma, s}, i) \\
\end{aligned}
\end{equation}
To deliver a spike from the source neuron $s$
to the MPI group
$\alpha \gets G_{\sigma, s, j}$,
its corresponding positions 
$i \gets Q_{\sigma, s, j}$
is sent by a collective call to the MPI group 
$\alpha$.
This position is used by each
target MPI process of the same group
to extract the local image neuron index,
$I_{\alpha, \sigma, \tau, i}$,
and thus to deliver the spike
through the connections outgoing from this node.

\subsubsection{{\it Random, fixed in-degree} connection rule
for neuron populations distributed across multiple MPI processes}
The \texttt{RemoteConnect} method described previously requires specifying the MPI process
$\sigma$ where the source neurons are located and the MPI process $\tau$ where the target neurons are located.
However, in many cases, it is useful to represent populations of neurons
distributed across multiple MPI processes, and apply probabilistic connection rules
to these populations as a whole.
This is the case of the scalable balanced network model used in this work
and described later.
To this end, we can define a distributed population
of source neurons, 
$\mathcal{S}$,
as a collection of $K$ subpopulations
$\mathbf{s}^{(0)}, ..., \mathbf{s}^{(K-1)}$
allocated in $K$ MPI processes, of rank
$\sigma_0, ..., \sigma_{K-1}$
\begin{equation}
\mathcal{S} = \{ ( \sigma_0, \mathbf{s}^{(0)}), ...,
( \sigma_{K-1}, \mathbf{s}^{(K-1)} ) \}
\end{equation}
and similarly a distributed population
of target neurons,
$\mathcal{T}$,
as a collection of $H$ subpopulations
$\mathbf{t}^{(0)}, ..., \mathbf{t}^{(H-1)}$ 
allocated in $H$ MPI processes, of rank
$\tau_0, ..., \tau_{H-1}$
\begin{equation}
\mathcal{T} = \{ ( \tau_0, \mathbf{t}^{(0)} ), ...,
( \tau_{H-1}, \mathbf{t}^{(H-1)} ) \}
\end{equation}
The connection rule {\it random, fixed in-degree with multapses}~\cite{Senk2022}
with $K_\text{in} = c$
prescribes that for each neuron in the target population,
$c$ connections are created by randomly extracting the source neuron of each of these connections with a uniform distribution from all neurons in the source population.
Since in the approach used in this work, connections are created in the MPI process where the target neuron is located, this operation can be performed independently for each process $\tau_j$, randomly extracting, for each target neuron of this process,
$t^{(j)}_i$,
$c$ elements
of the set $\mathcal{S}$.
This procedure produces
$c \times N^{(j)}_\mathbf{target}$ triplets,
where $N^{(j)}_\mathbf{target}$ is the number of neurons in the
$j$-th target subpopulation:
\begin{equation}
\mathcal{C} = (
(\widetilde{\sigma}_0, \widetilde{s}_0, t^{(j)}_0), ..., 
(\widetilde{\sigma}_{c N^{(j)}_\mathbf{target} - 1}, 
\widetilde{s}_{c N^{(j)}_\mathbf{target} - 1}, 
t^{(j)}_{N^{(j)}_\mathbf{target} - 1} )
\end{equation}
These triplets are sorted using as a first sort key
$\widetilde{\sigma}_i$ and as a second $\widetilde{s}_i$.
\begin{equation}
\mathbf{C} = \mathbf{sort}(\mathcal{C})
\end{equation}
and the resulting sequence,
$\mathbf{C}$ is divided
into distinct subsequences based on the value of the source process,
$\sigma$.
For each of these subsequences, the
\texttt{RemoteConnect} method described previously is then invoked,
using as an argument the 
$\sigma$ value associated with the subsequence
to specify the source MPI process,
and using a special connection rule,
{\it assigned-nodes}, which, instead of generating
the indexes of the source and target neurons of the connections,
uses already assigned values, in this case the values of $s$ and $t$
of the elements of the subsequence.

\subsubsection{GPU memory levels}\label{sec:opt_levels}
Four GPU memory levels (GMLs) have been implemented to balance the GPU memory occupation and the time-to-solution. Indeed, when performing large-scale simulations of networks with natural neuron and connection density, a significant part of the GPU memory is occupied by data structures for mapping remote connectivity, and some of them may be stored in CPU memory to aim for GPU memory optimization. These GPU memory levels can be employed both with point-to-point and collective communication approaches. The levels, ordered by increasing GPU memory usage, are designed as follows:
\begin{itemize}
\item \textbf{GPU memory level $0$}: The maps of the remote source neurons, the maps to their local-images, the first index, and the number of the outgoing connections of each remote neuron are stored in CPU memory.
Note that the connections are sorted based on the index of the source neuron as the first sorting key~\cite{Golosio23_9598}, consequently to identify all the connections outgoing from a given neuron, it is sufficient to indicate the index of the first of these connections and their number.
\item \textbf{GPU memory level $1$}: The maps of the remote source neurons to their local images, the first index, and the number of the outgoing connections of each remote neuron are stored in CPU memory. From this level on, it is assumed that all source nodes specified as arguments to a \texttt{RemoteConnect} call should have an image in the target MPI process, without checking if they are actually used in at least one connection in such a process. This makes the remote connection creation faster. However, if a significant fraction of the source nodes is not actually used in any connection, this leads to a waste of GPU memory.
For some connection rules, such as \textit{one-to-one}, \textit{all-to-all} and \textit{fixed-in-degree}, it is guaranteed that each source neuron of the set passed as argument of a call to the Connect method is actually used in at least one connection. On the other hand, there are rules, such as \textit{fixed-in-degree} and \textit{fixed-total-number}, for which it is not necessarily true that every source neuron given as an argument to the call to the Connect method is actually used in some connections. 

Creating an image neuron for each source neuron of the set given as argument of the \texttt{RemoteConnect} command without checking whether or not a connection is used induces a slight waste of GPU memory, which in the case of a random network is relatively small when the number of MPI processes is smaller than the average number of outgoing connections for each neuron, but starts to become relevant when the number of MPI processes is comparable or larger than this number.
\item \textbf{GPU memory level $2$}: The maps of the remote source neurons to their local images and the first index of the outgoing connections of each remote neuron are stored in GPU memory. The number of outgoing connections of a remote neuron is computed on the fly when needed.
\item \textbf{GPU memory level $3$}: The maps of the remote source neurons to their local images, the first index, and the number of the outgoing connections of each remote neuron are stored in GPU memory.
\end{itemize}

The GPU memory level $2$ is used as the default for simulation in NEST GPU.

\subsection{Models employed for performance evaluation}\label{sec:models}
\subsubsection{Multi-Area Model}\label{sec:MAM}
The Multi-Area Model (MAM) is a large-scale spiking model representing 32 areas of the macaque vision-related cortex \cite{Schmidt2017, Schmidt2018}, each one representing a cortex volume obtained by extruding $1$\,mm$^2$ patch through the full cortex thickness. The areas are chosen among those that have visual function or have strong connections with the latter, and are designed with a laminar structure, with layers 2/3, 4, 5, and 6 containing an excitatory and an inhibitory population each, with an intra-area connectivity similar to that of the cortical microcircuit of~\cite{Potjans_2014}. The only exception is area TH, which lacks layer 4. The inter-area connectivity is based on axonal tracing and quantitative tracing data, and the neuron densities per area is determined from empirical measurements, cytoarchitectural type definitions of areas, and the thicknesses of the cortical layers. The firing rates of the populations are adjusted through a mean-field approach~\cite{Schuecker2017}, and the initial membrane potentials of the neurons are normally distributed as in~\cite{Schmidt2018}.

Based on the synaptic strength of the cortico-cortical synapses, the model can fall into different activity states. When the synaptic strengths are equal between local and cortico-cortical synapses, the model shows a stationary \textit{ground state} of activity, characterized by low activity fluctuations and limited interactions between neurons of different areas. As the cortico-cortical synaptic strength increases, the model shows irregular activity with higher rates and enhanced inter-area interactions, named \textit{metastable state}. Further details on the model may be found in the original publications~\cite{Schmidt2017, Schmidt2018}. Such a model is simulated using the point-to-point communication approach. To eliminate the effects of transient dynamics at the beginning of the simulation and to assess the performance in more stable conditions, each simulation was preceded by a warm-up of $500$\,ms of model time, followed by a simulation model time of $10,000$\,ms. The simulation time resolution was $0.1$\,ms.
The MAM simulations used the default GPU memory level (i.e., level 2) to balance between memory occupation and time-to-solution.

\paragraph{Area packing}\label{sec:area_packing}
In our initial study with NEST GPU~\citep{Tiddia2022}, the MAM was implemented so that each area is contained within a single GPU.
This leveraged the fact that intra-area connectivity is higher than inter-area connectivity, hence the biggest spike communication payload of each area remained contained within each GPU device and lowered the total bandwidth required for each MPI communication round.
At the time, the GPUs used for the study were the NVIDIA V100s (see Section \ref{sec:hardware}), which also limited the number of simultaneously instantiated areas within a single GPU.
With larger GPUs available, such as the NVIDIA A100s (see Section \ref{sec:hardware}), it stands within reason to explore different area distribution schemes to study whether the increase in neuron network density per GPU and consequently the decrease of MPI communication would enable better performance.
As a first network distribution scheme, we implemented an ``\textit{area packing}'' algorithm to distribute multiple areas within a single GPU while balancing the load across each GPU.
The algorithm is based on the classic ``0-1 Knapsack problem''~\cite{Knapsack1990} where each area can only be assigned once and the area weight is measured by the sum of the total number of incoming connections to the area and the number of neurons within the area.
As the implementation relies on knowing the connectivity information of the model before instantiating neurons and connections, the \textit{area packing} algorithm is run during the initialization of the model script (through the Python interface) and scans the connectivity data files included in the model implementation.

\subsubsection{Scalable balanced network} \label{sec:scalable_net}
The scalable balance network was closely modeled after the ``HPC benchmark'' from NEST CPU, historically used as a performance reference for computing clusters. Unlike the original CPU-based implementation (see~\cite{Helias2012,Kunkel2014}), the GPU version of the network model instantiates static excitatory-to-excitatory connections. The average firing rate of excitatory and inhibitory populations results below $10$\,Hz and does not depend on the network size.
In line with the CPU-based implementation, the GPU version also employs collective MPI communication, since densely connected large-scale simulations strongly benefit from using this protocol for the exchange of spikes among the MPI processes. A warm-up time of $500$\,ms was simulated to discard transients, with the following simulation time being of $1,000$\,ms. The simulation time resolution was $0.1$\,ms.

\paragraph{Details of scalable architecture}
In the model implementation in NEST GPU, the number of incoming connections per neuron is fixed to $K_{in}=11,250$, with $K_{in, E} = 9,000$ excitatory and $K_{in, I} = 2,500$ inhibitory connections, whereas the number of neurons on each MPI process depends on a scaling parameter so that the model has $11,250*\texttt{scale}$ neurons, with $9,000*\texttt{scale}$ excitatory and $2,250*\texttt{scale}$ inhibitory neurons.

Thus, the total network size is not only determined by the \texttt{scale} parameter, but also by the number of processes used. As opposed to the CPU implementation, where the \texttt{scale} parameter determines the total network size, regardless of the number of processes used in the computing system.

\subsection{Simulation phases and optimization}\label{sec:benchmarking}
The time-to-solution of the simulation can be divided into different phases: \textit{network construction} and \textit{simulation} (or \textit{state propagation}). The former comprises all the steps needed before the simulation loop begins, and it can be further divided as follows:

\begin{itemize}
\item Initialization, i.e., the time needed for preparing the model and the simulator, importing the libraries and the classes needed to perform the simulation;
\item Neuron and device creation, which instantiates neurons and devices, such as spike generators;
\item Neuron and device connection. This can be divided into
\begin{itemize}
    \item local connection, which handles the connection among neurons and devices belonging to the same MPI process;
    \item remote connections, which handles the connection among neurons simulated in different MPI processes;
\end{itemize}
\item Simulation preparation, needed to organize the connections and initialize data structures for spike delivery.
\end{itemize}

The time taken by network construction represents an overhead that is less relevant for simulations with relatively long biological simulated times, but can become a significant bottleneck for shorter model times, particularly when a large number of simulations need to be run, e.g., for model parameter optimization.

After network construction is completed, the simulation loop can start. The time needed to complete the simulation loop is measured through the \textit{real-time factor} ($\text{RTF}$), defined as the ratio between the wall clock time ($T_\text{wall}$) and the biological time simulated ($T_\text{model}$):

\begin{equation}
\label{eq:rtf}
    \text{RTF} = \frac{T_\text{wall}}{T_\text{model}}
\end{equation}

When the simulations are performed for benchmarking purposes, the recording of spikes or any variables from neurons and synapses is disabled. Moreover, times are measured at the Python interface level to be agnostic to the implementation.

Another key concept for the optimization of SNN simulations on GPUs is memory usage. GPUs necessarily need to interact with the CPU, so in general, simulations will use both CPU and GPU memory. However, in GPU-based approaches, CPU memory is typically underutilized, partly because most resources are allocated to the GPU and CPU memory is typically much larger even in GPU-equipped cluster nodes. Therefore, the limiting factor in these systems is GPU memory. Typically, in models of neuroscience interest, the number of connections is several orders of magnitude greater than the number of neurons (in the cerebral cortex, typically on the order of $10^3-10^4$, while in the cerebellum, even $10^5$), so the memory required for connection parameters is much greater than that required for neurons.
Appendix \ref{app:indegree_scaling} assesses the scaling performance across biologically plausible in-degree ranges. This analysis demonstrates that our approach is ultimately limited by the total number of synapses fitting into memory and not by the number of synapses per neuron.
In addition, a significant amount of memory may be required to store the previously mentioned structures used to efficiently communicate and distribute spikes between neurons located in different MPI processes. When simulating large networks, it is essential to monitor how the memory used by these structures scales with the size of the network.

\subsection{Validation of spiking statistics}\label{sec:validation_spikes}
The novel approach for network construction on multi-GPU systems leads to a change in the random numbers in the probabilistic connection rules. Even if the same seed is used, the resulting network instance is different than in the previous version of the simulation code, and spiking activity can be compared only on a statistical level. Thus, it is needed to verify that the network dynamics of the models simulated are qualitatively preserved.

To validate the new network construction method, we performed simulations of the MAM, comparing the results of the simulations obtained using the two versions of our reference implementation. Indeed, \textit{offboard} version has already been validated in~\cite{Tiddia2022} in the simulation of the MAM taking the CPU version of NEST as a reference. In this work, we employ the same spike statistics validation protocol, extracting three statistical distributions of the spiking activity for each population of the model:

\begin{itemize}
    \item Time-averaged firing rate for each neuron;
    \item Coefficient of variation of inter-spike intervals (CV ISI);
    \item Pairwise Pearson correlation of the spike trains obtained from a subset of 200 neurons for each population.
\end{itemize}

The distributions obtained using the different network construction methods are then compared as described in detail in Appendix \ref{app:validation_spikes}.

\subsection{Hardware}\label{sec:hardware}
We performed the simulations on two data centers, with different GPU configurations. In particular, we used JUSUF~\cite{VonStVieth2021} compute cluster from the Jülich Supercomputing Centre, and the Leonardo Booster \cite{Turisini2024} from the CINECA supercomputing Centre. The following shows the specifications of these two systems on a per-node basis:

\begin{itemize}
    \item JUSUF: 2 $\times$ AMD EPYC 7742, 2× 64 cores, 2.25 GHz; NVIDIA V100, 1530 MHz, 16 GB HBM2e, 5120 CUDA cores
    \item Leonardo Booster:  1 $\times$ Intel Xeon Platinum 8358 CPU, 32 cores, 2.6 GHz (Icelake); 4 $\times$ NVIDIA custom Ampere A100 GPU, 64 GB HBM2, 6912 CUDA cores
\end{itemize}

JUSUF uses CUDA version 12.0, whereas Leonardo Booster uses CUDA version 12.1. Moreover, the NVIDIA A100 GPUs of the Leonardo Booster show a GPU memory that is four times higher than that of the JUSUF GPU cards. This is relevant especially for strong scaling experiments, since each GPU can handle the simulation of a larger network.

\section*{Funding}
This project has received funding from
the European Union’s Horizon 2020 Framework Programme for Research and Innovation under Specific Grant Agreement No. 945539 (Human Brain Project SGA3),
the European Union’s Horizon Europe Programme under the Specific Grant Agreement No. 101147319 (EBRAINS 2.0 Project), the Italian Ministry of University and Research (MUR) PNRR project FAIR PE0000013- CUP I53C22001400006, funded by NextGenerationEU, MUR PNRR project e.INS Ecosystem of Innovation for Next Generation Sardinia – spoke 10 - CUP F53C22000430001 – MUR code: ECS00000038, the BRAINSTAIN project, funded by INFN-CSN5, and the Juelich Research Centre intramural STEF fund for the update of instruments.

\section*{Acknowledgments}
The authors gratefully acknowledge the use of Fenix Infrastructure resources, which are partially funded from the European Union’s Horizon 2020 research and innovation programme through the ICEI project under the Grant Agreement No. 800858.
We acknowledge the CINECA award under the ISCRA initiative, for the availability of high-performance computing resources and support.
The authors are also grateful to Daniele di Bari for his support in the usage of the CINECA computing infrastructure.

\section*{Code Availability Statement}
The data that support the findings of this study are openly available at the following URL/DOI: \url{https://doi.org/10.5281/zenodo.17627864}.

\bibliography{bibliography}

\appendix

\section{Validation details}\label{app:validation_spikes}
As described in Section \ref{sec:validation_spikes}, the new network construction method needs a validation protocol to ensure that models created using probabilistic connections and having the same parameters show the same network dynamics from a statistical point of view. In order to achieve this, we collect the spiking activity of every neuron population of the Multi-Area Model of~\cite{Schmidt2017}, computing for each of them three distributions: the average firing rate of the populations, the coefficient of variation of inter-spike intervals (CV ISI), and the pairwise Pearson correlation of the spike trains for each population.

Every simulation has a pre-simulation phase of $500$\,ms followed by a simulation of $60$\,s of network dynamics. The pre-simulation phase is needed to avoid spiking activity transients that naturally occur at the simulation's beginning. In contrast, the simulation phase simulates $60$\,s of neural activity in order to let the activity statistic converge. Indeed,~\cite{Dasbach2021} shows that relatively long network dynamics is needed to distinguish the actual activity statistics of a network from a random process.

\begin{figure}[H]
    \centering
    \includegraphics[width=\columnwidth]{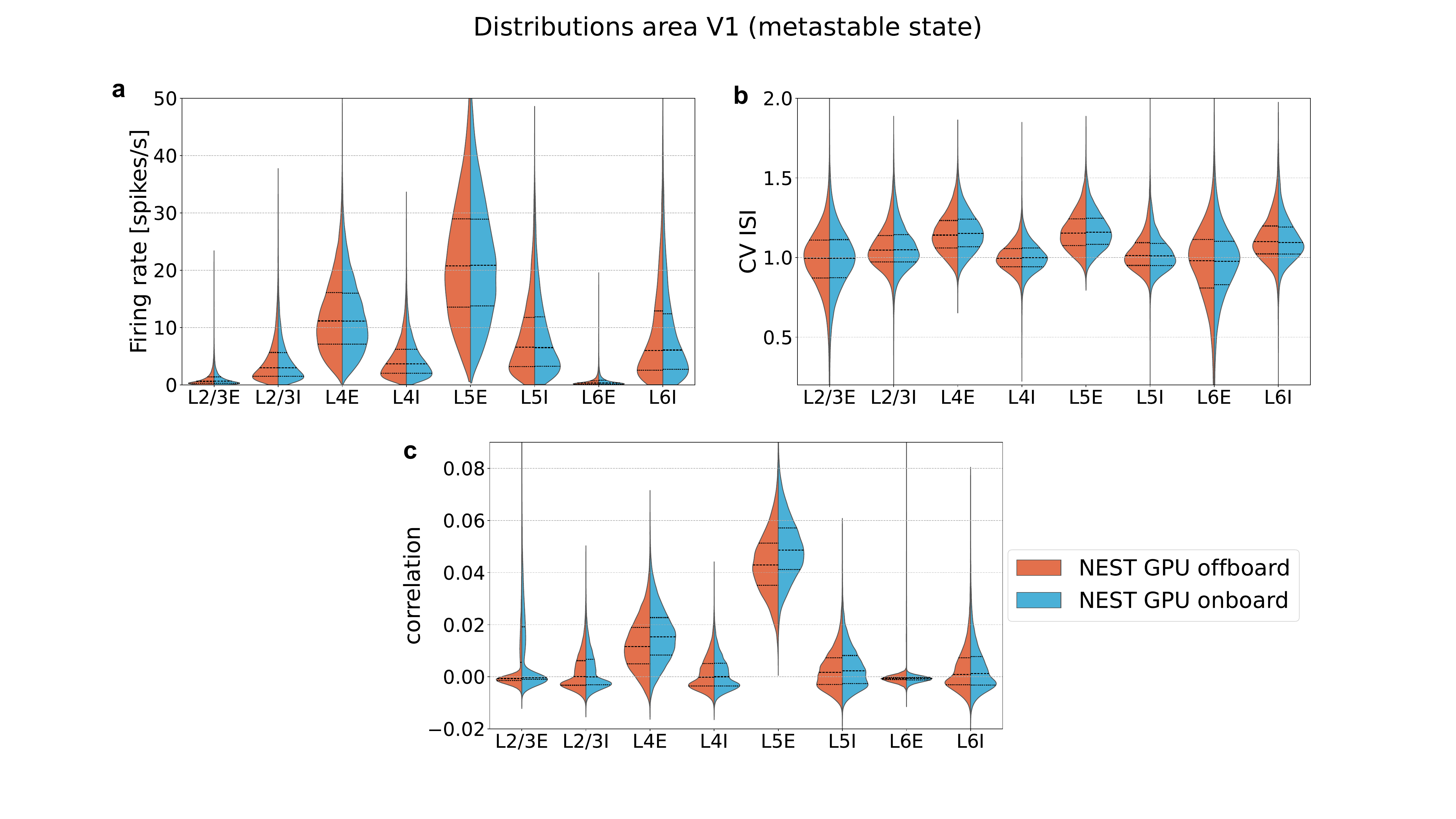}
    \caption{Violin plots of the distributions of firing rate (\textbf{a}), CV ISI (\textbf{b}), and Pearson correlation (\textbf{c}) for a simulation for the populations of the cortical microcircuit model with (sky blue distributions, right) or without (orange distributions, left) the new method for network construction. The distributions are obtained with the \texttt{seaborn.violinplot} function of the Seaborn~\cite{seaborn} Python library, which smooths the distributions using the Kernel Density Estimation method~\cite{Rosenblatt1956, Parzen1962} with Gaussian kernel and with bandwidth optimized through the Silverman method~\cite{silverman86}.}
    \label{fig:distribV1}
\end{figure}

Only the spikes emitted during the latter phase are recorded and employed to obtain the aforementioned distributions. Figure \ref{fig:distribV1} depicts the violin plots of the distributions obtained for one of the $32$ areas of the model in the metastable state. As shown in~\cite{Tiddia2022}, such a state of the network dynamics is characterized by a high degree of variability even when performing multiple simulations on the same simulation code using different seeds for random number generation.

The distributions obtained by the two versions of the library are mostly similar, however, some differences may occur. The differences that arise between two distributions can be quantitatively estimated using the Earth Mover Distance (EMD) metric, which is equivalent to the first-order Wasserstein distance and can be computed using the \texttt{scipy.stats.wasserstein\_distance} of the SciPy Python library~\cite{SciPy}. We then followed the validation protocol described in~\cite{Tiddia2022}, which aims to evaluate the degree of fluctuations that arise between the statistical distribution when simulations are performed through the same simulation code using different seeds and the ones between simulations performed using the two different simulation codes. If the latter fluctuation is compatible with the former, it indicates that the novel methods implemented in the simulation code do not add variability to the statistical results. 

The degree of the fluctuation is measured through EMD for every statistical distribution and every population of the model. For a better comparison, three sets of simulations have to be performed using different seeds for random number generation: two sets of simulations with the \textit{offboard} version, and a set of simulations with the \textit{onboard} version, that implements the novel construction method. With this, we can compare the simulations performed with the \textit{offboard} version to evaluate the fluctuations that arise from the change of the seed, whereas one of the sets of simulations just mentioned is compared with the set of \textit{onboard} simulations to estimate the fluctuation between the simulation codes with different network construction methods. The comparisons are made in a pairwise fashion, hence, for each population and statistical distribution, we collect an amount of values of EMD equal to the number of simulations performed for each set.

In this case, every set of simulations is composed of $20$ simulations. This number is the result of a compromise between the need for more data to evaluate the fluctuations measured with the EMD and the amount of compute time and storage needed to perform the simulations required. For instance, three sets of $20$ simulations of one minute of spiking activity recorded in the metastable state of the MAM need up to $20$\,TB of memory. Moreover, the compute time needed, especially for the simulations with the previous method for network construction, is an additional limiting factor.

\begin{figure}[H]
    \centering
    \includegraphics[width=\columnwidth]{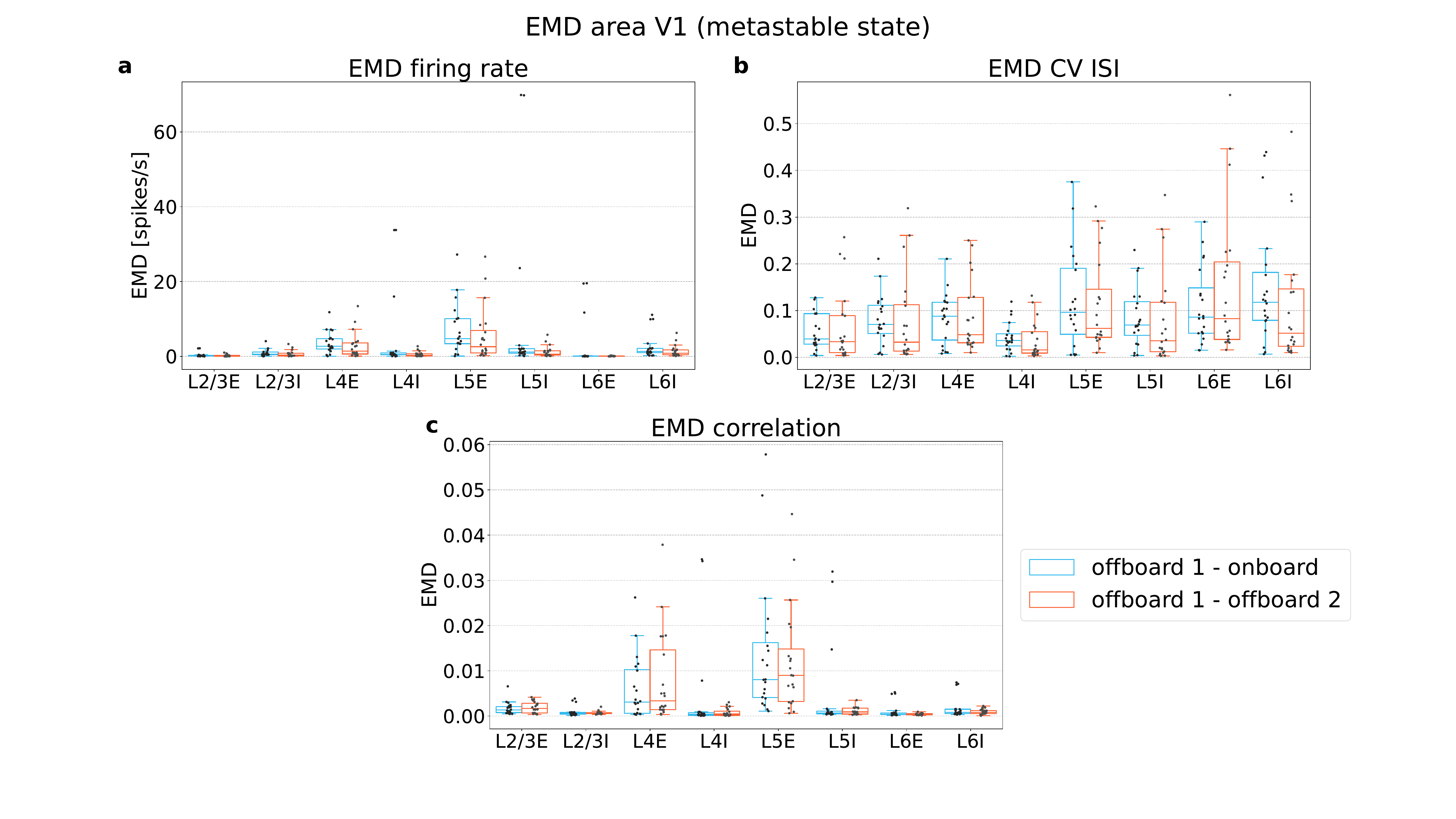}
    \caption{Box plots of the Earth Mover’s Distance comparing side by side firing rate (\textbf{a}), CV ISI (\textbf{b}) and Pearson correlation (\textbf{c}) of the two versions of the code (sky blue boxes, left) and the previous version using different seeds (orange boxes, right). The central line of the box plot represents the median of the distribution, whereas the extension of the boxes is determined by the interquartile range of the distribution formed by the values of EMD of each comparison. The whiskers show the rest of the distribution as a function of the interquartile range, and the dots represent the $20$ values of the EMD from which the boxes are generated.}
    \label{fig:EMDV1}
\end{figure}

Figure \ref{fig:EMDV1} shows the values of EMD for the three statistical distributions for the first area of the MAM.

The box plots of Figure \ref{fig:EMDV1} show a similar distribution of the values of EMD for the comparisons performed, which leads to the conclusion that the novel method for network construction does not add variability to the statistical distribution with respect to the previous version of the library.

\section{MAM area packing results}\label{app:mam_area_packing}

\begin{figure}[H]
    \centering
    \includegraphics[width=\linewidth]{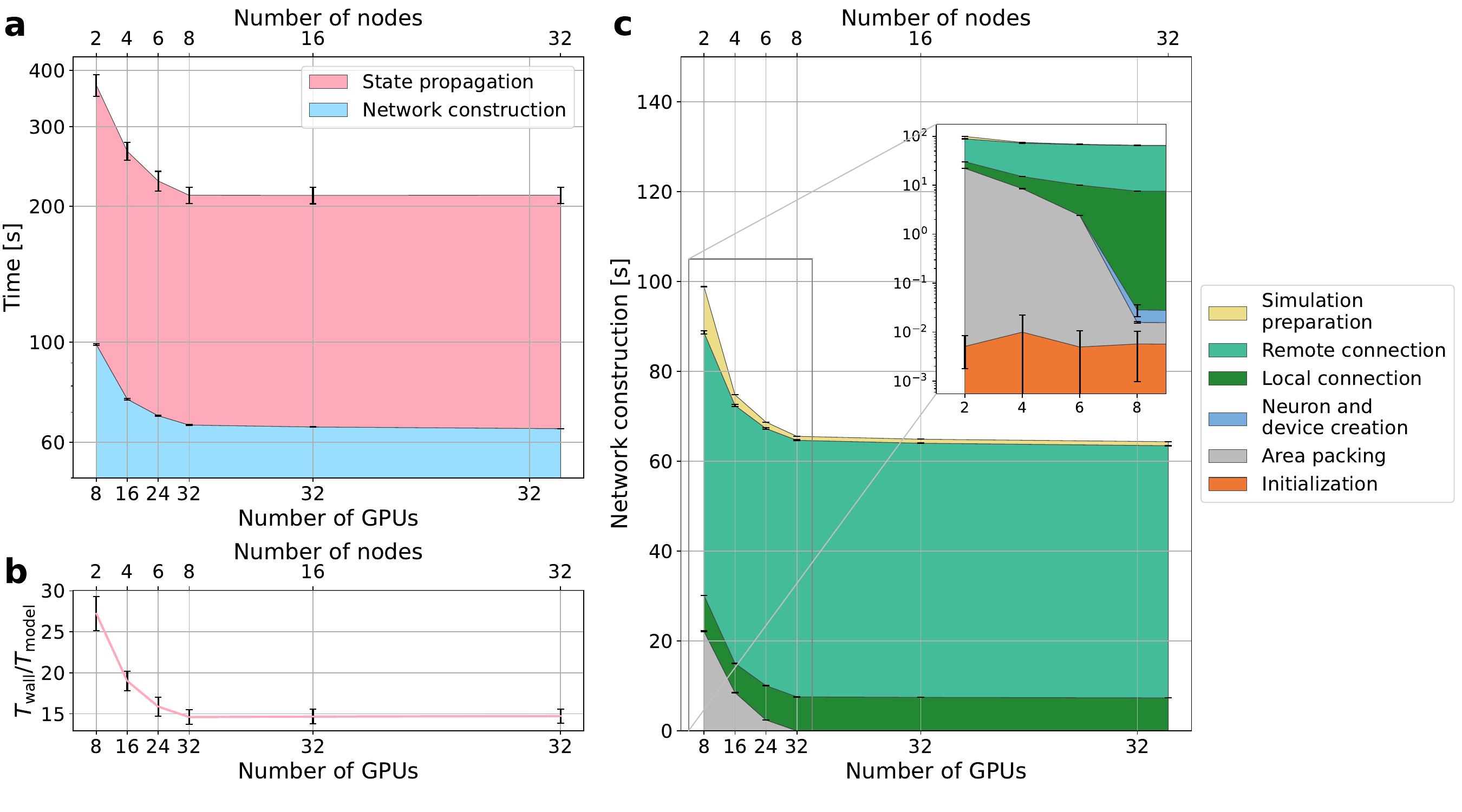}
    \caption{Performance of the \textit{onboard} version on simulation of MAM metastable state leveraging the area packing algorithm on Leonardo Booster. Mean data shown from averaging over $10$ simulations using different random seeds, black error bars represent the standard deviations.
    For the $2–8$ node configurations, four GPUs per node were utilized. For the $16$-node configurations, two GPUs per node were used, and for the $32$-node configuration, one GPU per node was used. All configurations use $1$ MPI process per GPU. (\textbf{a}) Absolute wall-clock time for the network construction and state propagation for a biological model time $T_\text{model} = 10$\,s.
    (\textbf{b})\,State propagation described as real-time factor. (\textbf{c}) Network construction time divided into subtasks.}
    \label{fig:MAM_area_packing}
\end{figure}

The area packing algorithm described in Section \ref{sec:area_packing} was developed to fit more areas of the MAM model inside the same GPU to enable GPU memory-optimized simulations of the MAM. To this aim, the NVIDIA A100 GPUs of the Leonardo Booster supercomputer~\cite{Turisini2024}, with $64$\,GB of GPU memory, can be used to test the performance of the MAM when this algorithm is employed. Indeed, the V100 GPUs of the JUSUF cluster, with $16$\,GB of GPU memory, could fit up to a model area per GPU. Figure~\ref{fig:MAM_area_packing} shows the performance benchmarks of the MAM when area packing is employed.

The algorithm enables the MAM to be executed on two compute nodes with $4$ GPUs each (i.e., $2$ nodes), even if with a longer time-to-solution with respect to the implementations in which the model areas are simulated into a larger number of GPUs.
Figure~\ref{fig:MAM_area_packing}\textbf{b} depicts the real-time factor, which aligns to the one shown in Figure~\ref{fig:MAM_simulations}\textbf{b} when the same number of GPUs is employed. Indeed, despite different GPUs are used in the two cases, the real-time factor of the simulation can be similar with these devices, as also shown in \cite{Golosio23_9598} for the simulation of the model used as a building block for the MAM.
Figure~\ref{fig:MAM_area_packing}\textbf{c} shows that the increased network construction time for few MPI processes is mainly due to the increased load on the area packing algorithm. 
The plateau occurs at eight nodes, thus in a condition in which $32$ GPUs are used. Indeed, this is expected since the difference between the last three configurations is only the number of GPUs used on each compute node.

Although NVLink 3.0 (used for intra-node GPU communications~\cite{Turisini2024}) offers higher bandwidth than InfiniBand, no major performance gain was observed, as inter-node communication remains the limiting factor. An improvement in this regard could be achieved by simulating the entire model in a single compute node, which is not possible with the memory limits of the GPUs used here. Besides, one could explore more specialized communication schemes such as the NVIDIA Collective Communications Library (NCCL) for GPUs if suitable high-speed interconnects within a node or NVIDIA networking across nodes is available (\url{https://developer.nvidia.com/nccl}, last visited: 24.10.2025).

\section{Scalable balanced network simulations using different network scale parameter}\label{app:different_scales}

\begin{figure}[H]
    \centering
    \includegraphics[width=0.7\linewidth]{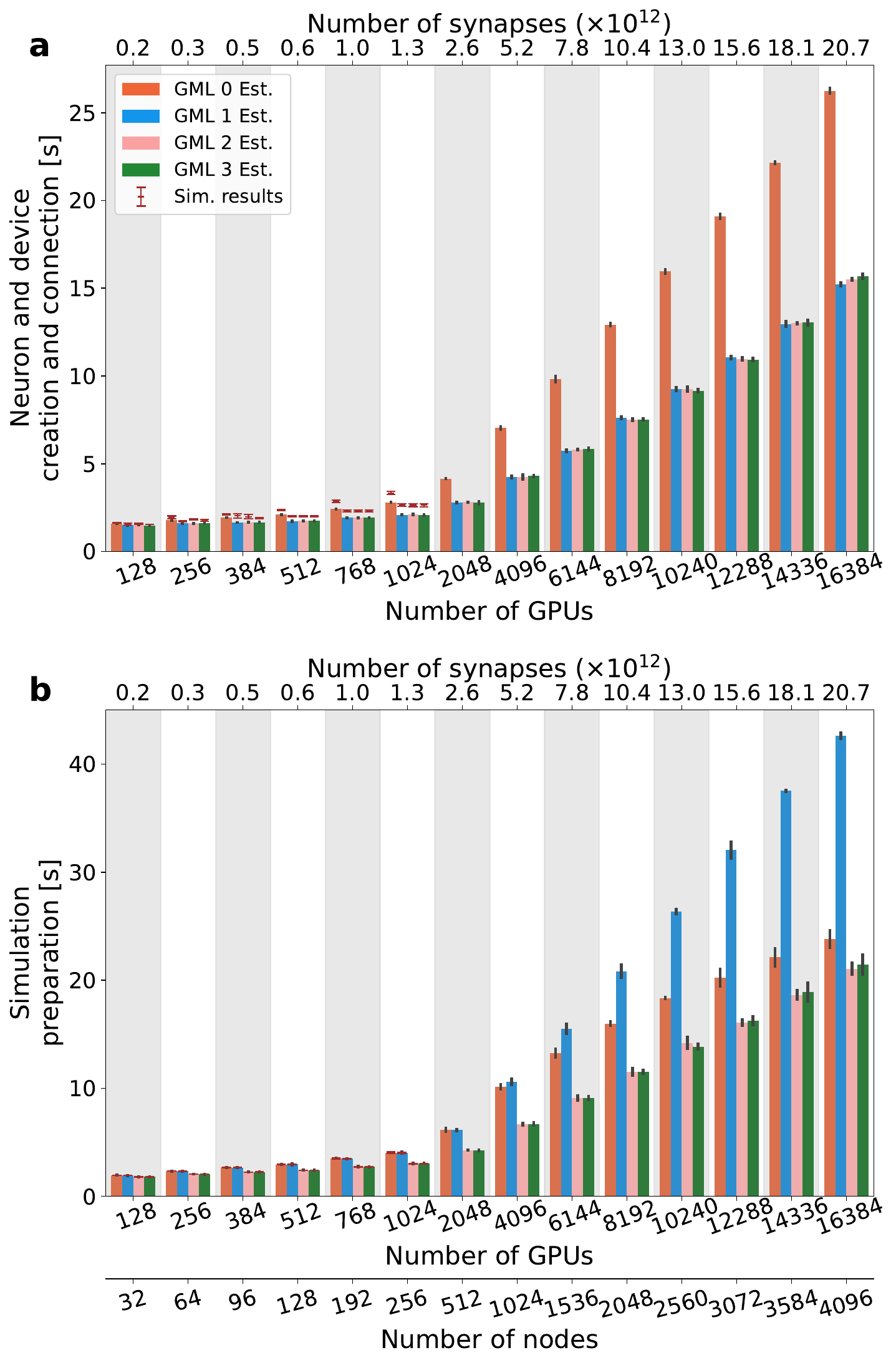}
    \caption{Network construction time divided into neuron creation and connection (\textbf{a}) and simulation preparation (\textbf{b}) times of the scalable balanced network model simulated at scale $10$ evaluated across all MPI processes as a function of the number of cluster nodes for the three GPU memory levels. The bars represent an estimation using four MPI processes on a single node.
    The horizontal line markers show the time averaged across the MPI processes using the GPU memory levels. Errors for bars and horizontal line markers represent their standard deviation, with most of the values showing little variability.}
    \label{fig:-construct-and-preparation-time10}
\end{figure}

\begin{figure}[H]
    \centering
    \includegraphics[width=0.7\linewidth]{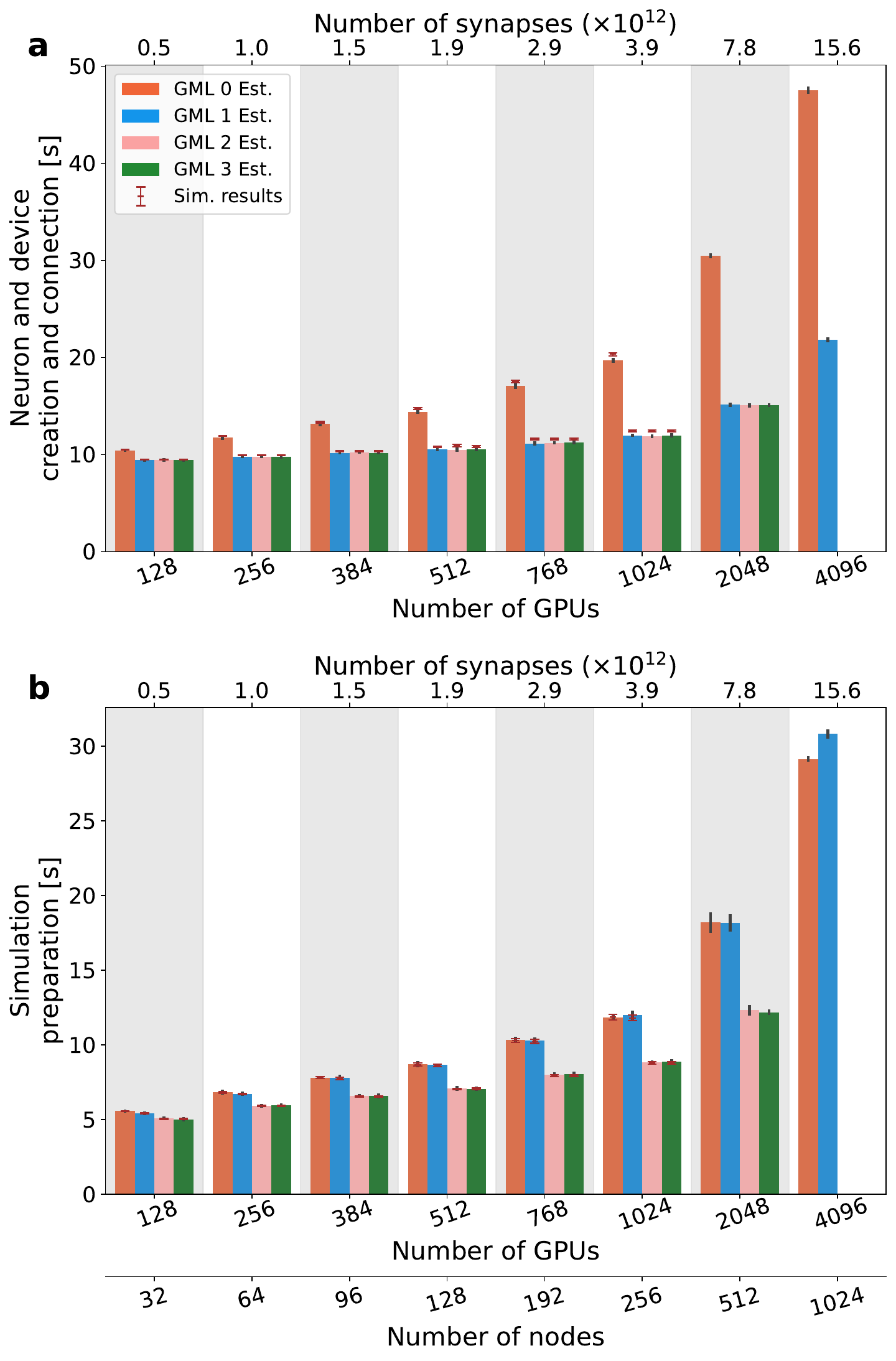}
    \caption{Network construction time divided into neuron creation and connection (\textbf{a}) and simulation preparation (\textbf{b}) times of the scalable balanced network model simulated at scale $30$ evaluated across all MPI processes as a function of the number of cluster nodes for the three GPU memory levels. The bars represent an estimation using four MPI processes on a single node.
    The horizontal line markers show the time averaged across the MPI processes using the GPU memory levels. Errors for bars and horizontal line markers represent their standard deviation, with most of the values showing little variability.}
    \label{fig:-construct-and-preparation-time30}
\end{figure}

In the main body of the manuscript, we performed simulations of the scalable balanced network model
setting the scale parameter to $20$, i.e., $2.25 \times 10^5$ neurons and $2.53 \times 10^9$ synapses per GPU. However, different scales can be adopted to simulate larger or smaller networks inside each node. To provide more insight into the network construction performance of networks of different sizes, Figure~\ref{fig:-construct-and-preparation-time10} and Figure~\ref{fig:-construct-and-preparation-time30} show construction and simulation preparation time in case of networks of scale $10$ (i.e., $1.125 \times 10^5$ neurons and $1.26 \times 10^9$ synapses per GPU) and $30$ (i.e., $3.375 \times 10^5$ neurons and $3.80 \times 10^9$ synapses per GPU), respectively. As can be seen, the network size, expressed in terms of total synapses, differs from that of Figure~\ref{fig:-construct-and-preparation-time}.

\section{Scalable balanced network with different in-degree scale parameter}\label{app:indegree_scaling}

The scalable balanced network described in Section \ref{sec:scalable_net} shows a fixed number of incoming connections per neuron, with the \texttt{scale} parameters used to change the number of neurons simulated on each MPI process. This way, the network size increases, both in the total number of neurons and connections. However, a different approach may be employed to test the capabilities of the simulation code by simulating a network when the number of incoming connections per neuron changes. Neurons in certain regions of the brain, such as the cerebellum~\cite{Popa2018}, may show up to $2\times 10^5$ incoming connections, and similar in-degree levels are not explored in the simulation of the network models proposed in this work, and it would be of interest to show the performance of the simulation code in such a framework.

In this regard, we performed simulations on a modified version of the scalable balanced network model. We defined a \texttt{in-degree\_scale} parameter, so that each MPI process has a total of $9,000 \times \texttt{scale} / \texttt{in-degree\_scale}$ excitatory neurons and $2,250 \times \texttt{scale} / \texttt{in-degree\_scale}$ inhibitory neurons, with $\texttt{scale} = 10$ and a fixed in-degree per neuron equal to $K_\text{in} = 11,250 \times \texttt{in-degree\_scale}$. Moreover, to preserve the average firing rate of the model, the synaptic weights for each connection are adjusted with the \texttt{in-degree\_scale} factor, so that the product of the number of in-degrees and their synaptic weights remains constant. This way, the total number of connections per MPI process is the same as for the scaling experiments described in the Results, granting similar GPU memory consumption. Similarly, the firing rate does not show significant differences with respect to the scalable network model configurations employed in the rest of the work.

\begin{figure}[H]
    \centering
    \includegraphics[width=0.9\linewidth]{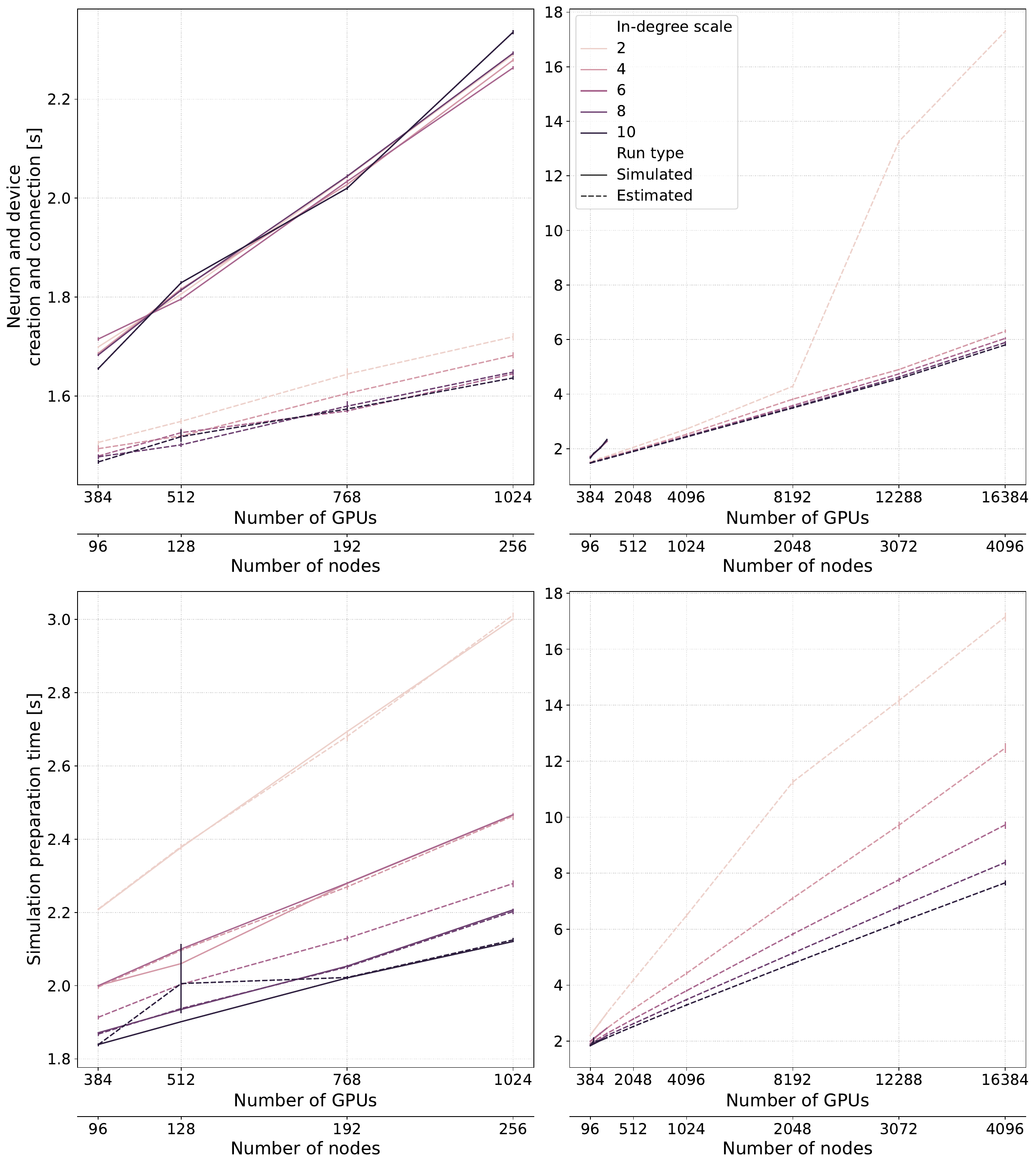}
    \caption{Neuron and device connection and creation and simulation preparation times of the scalable balanced network model for different values of \texttt{in-degree\_scale} as a function of the number of cluster nodes (four GPUs per node). Panels on the left represent the results in the range of compute nodes for which simulations are executed, whereas right panels extend to the range in which the performance is evaluated with the estimation method. Lines represent simulated results, whereas dotted lines represent estimated results. Vertical bars indicate the standard deviation.}
    \label{fig:indegree-test}
\end{figure}

This experiment has been performed using the GPU memory level $0$ since, as shown in the Results, it is the level capable of leveraging the largest number of compute nodes. Figure \ref{fig:indegree-test} shows neuron and device creation and simulation preparation times using different values for \texttt{in-degree\_scale}. The maximum value adopted for such a scaling parameter is $10$, i.e., the in-degree values range from $22500$ for $\texttt{in-degree\_scale} = 2$ to $112500$ for $\texttt{in-degree\_scale} = 10$.

The figure shows that both times diminish as soon as the \texttt{in-degree\_scale} increases. This is expected, since a larger parameter implies that the number of neurons diminishes, and thus also the number of image nodes, resulting in faster construction and sorting operations.

\section{Analysis of the difference between simulated and estimated values for the scalable balanced network model}\label{app:diff_est_sim}

In this appendix, we provide a detailed analysis of the discrepancies observed between estimated and simulated values of GPU memory consumption and during neuron and  device creation and connection times (Figures \ref{fig:gpu_memory} and \ref{fig:-construct-and-preparation-time}).

Regarding the divergence observed in Figure \ref{fig:gpu_memory} for the GPU memory peak, the construction of the network and of the maps for remote connectivity involves the use of temporary buffers in GPU memory, which are allocated in a non-deterministic manner due to the randomness of the connectivity operations. Therefore, the GPU memory usage peak shows slight variations (in the order of $1\%$) across different MPI processes, which is mostly observable in the case of GPU memory levels 2 and 3, where connection structures are explicitly stored and indexed in GPU memory.

This also affects the performance of the network construction; however, there are other sources of variability in the latter case, especially at the GPU memory level $0$. Large-scale distributed systems are subject to performance jitter, i.e., fluctuations in execution time caused by operating system activity and hardware-level variability. These effects accumulate with the number of nodes and are not captured by the estimation model. Additionally, during the simulations, the CPU thread responsible for connection construction is not bound to a specific core. Consequently, operating system scheduling may migrate the thread across cores, inducing cache invalidation, thus increasing variability in execution time.

\begin{figure}[H]
    \centering
    \includegraphics[width=\linewidth]{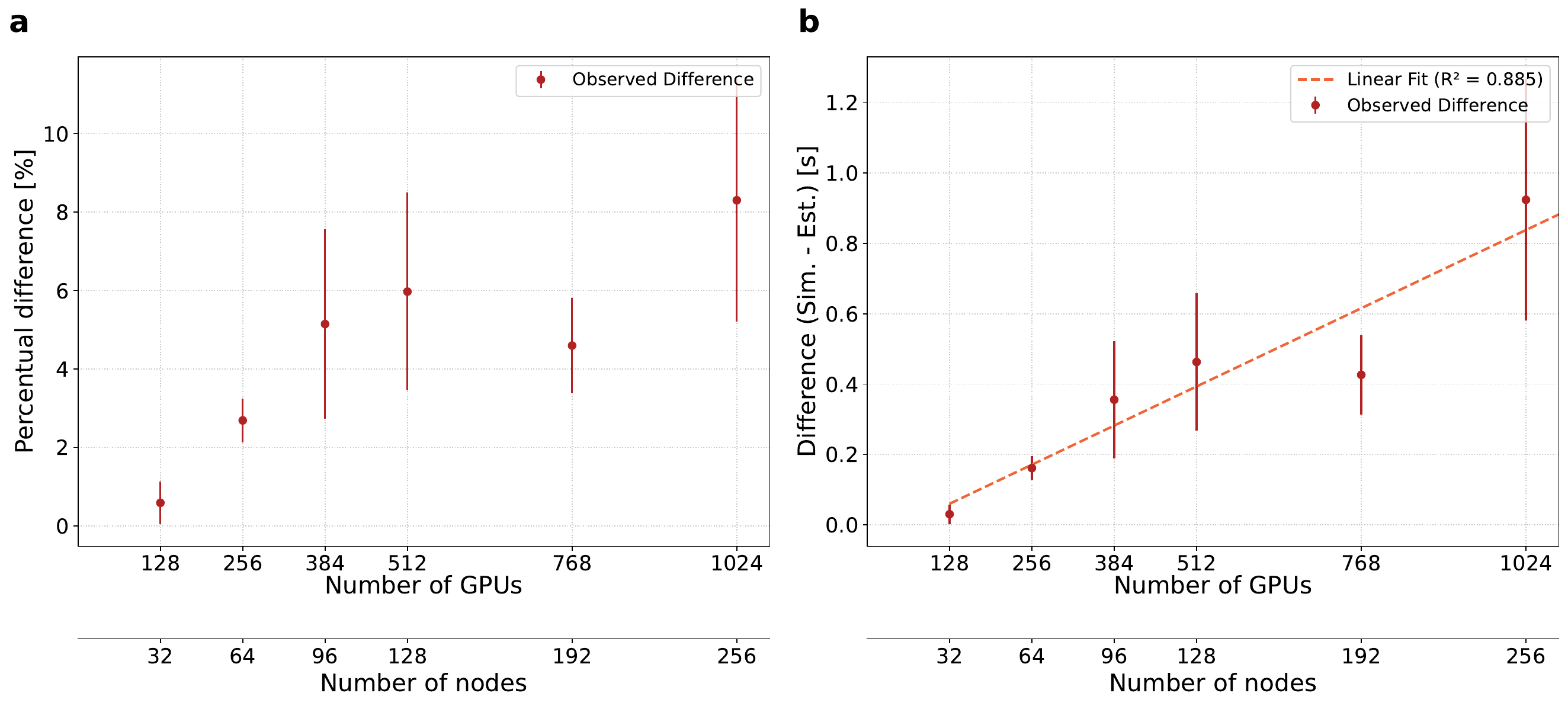}
    \caption{Difference between simulated and estimated neuron and device creation and connection time for GPU memory level $0$, measured as percentual difference \textbf{(a)} or absolute difference \textbf{(b)}. Dots represent the observed values, whereas the dotted line in panel \textbf{(b)} represents a linear fit of such differences. This data is derived from the simulations at the scalable network models using $\texttt{scale}=20$, thus using the data of Figure~\ref{fig:-construct-and-preparation-time}.}
    \label{fig:fit_difference}
\end{figure}

Differences between simulated and estimated network construction times in Figure \ref{fig:-construct-and-preparation-time} are shown in Figure \ref{fig:fit_difference}. Panel \ref{fig:fit_difference}a shows the percentage difference between simulated and estimated time, showing values for $256$ nodes below $10\%$. Panel \ref{fig:fit_difference}b shows the same discrepancy in absolute terms, and a linear fit has been performed to assess the degree of the discrepancies when thousands of compute nodes are employed for the simulation. Extrapolations of such behavior indicate discrepancies of about $14$\,s for $4096$ nodes.

\section{Example of remote spike communication through point-to-point MPI forwarding}\label{app:p2p}
\begin{figure}[H]
    \centering
    \includegraphics[width=1.1\linewidth]{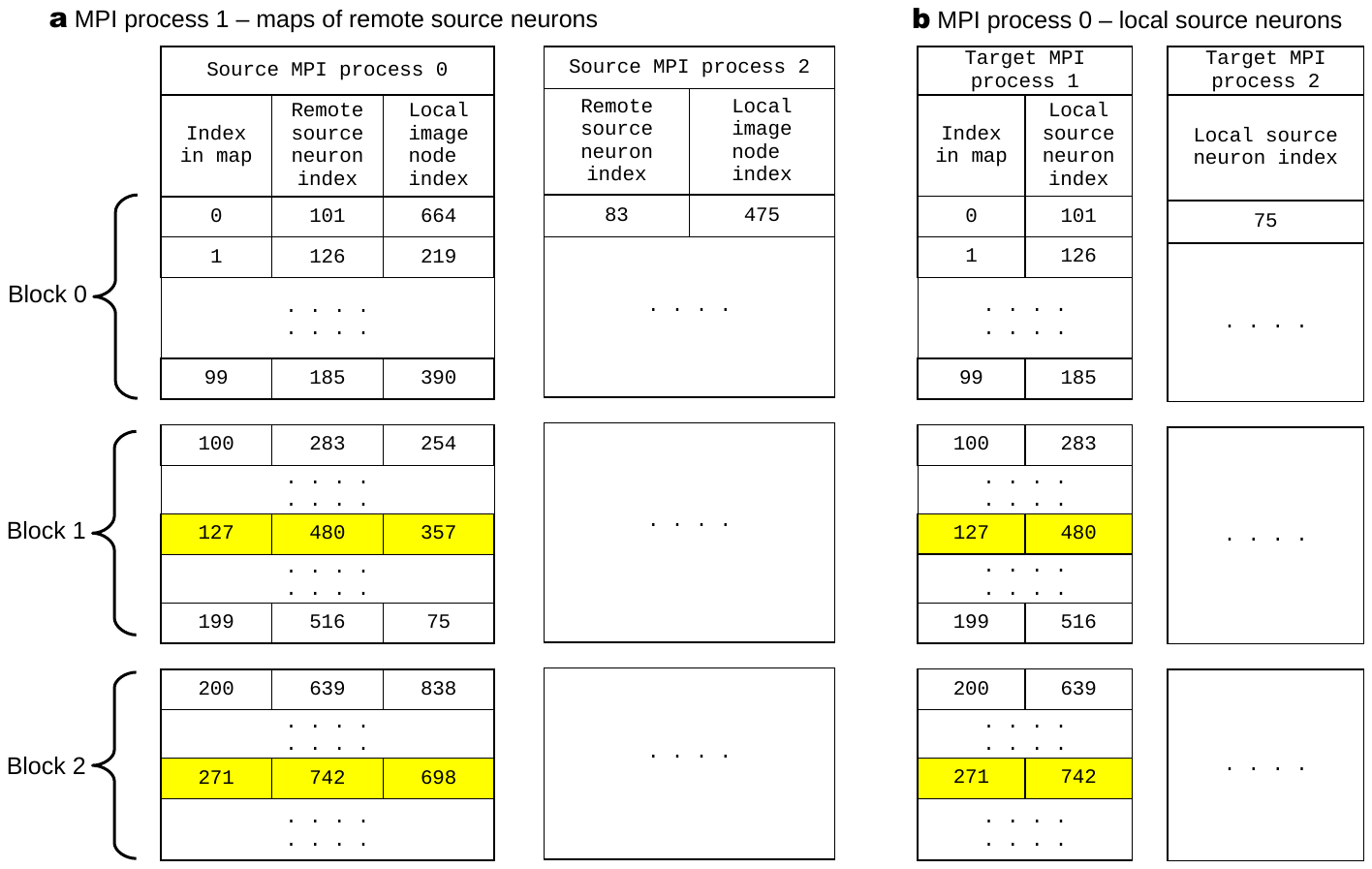}
    \caption{
    (\textbf{a}) 
organization of the maps
$(R_{\tau,\sigma, i},
L_{\tau,\sigma, i})$,
which associate 
remote source neurons to local image neuron indexes,
in the GPU memory of the target
MPI process, which has rank 
$\tau = 1$ in this example.
The left column represents the
position in the map, $i$,
while the other two columns
represent the
remote source neuron index,
$R_{\tau,\sigma, i}$
and the local index of its
image neuron
in the target MPI process,
$L_{\tau,\sigma, i}$.
(\textbf{b})
organization of the sequence
$S_{\sigma, \tau, i}$,
which represent the local source neuron index, in the GPU memory of the source
MPI process, which has rank
$\sigma = 0$ in this example.
The left column represents the
position in the sequence, $i$,
while the other column
represent the
local source neuron index,
$S_{\sigma, \tau, i}$.}
    \label{fig:p2p-maps}
\end{figure}

Figure
\ref{fig:p2p-maps}
\textbf{a}
illustrates how
the maps $(R_{\tau,\sigma, i},
L_{\tau,\sigma, i})$
are organized in the
GPU memory of an
MPI process, which has rank 
$\tau = 1$ in this example.
This process creates one map
for each possible source MPI process, i.e.
one for every MPI rank
excluding its own
(ranks $\sigma = 0$
and $\sigma = 2$
in this example).
These maps are organized
in fixed-size blocks
that are allocated dynamically
in GPU memory.

Figure \ref{fig:p2p-maps}
\textbf{b}
illustrates how
the sequences
$S_{\sigma, \tau, i}$
are organized in the
GPU memory of a different
MPI process, which has rank
$\sigma = 0$ in this example.
One sequence of this kind
is created
for each possible target MPI process, i.e.
one for every MPI rank
excluding that of the source
process itself
(ranks $\tau = 1$
and $\tau = 2$
in this example).
Note that we chose to show two different MPI processes in the two panels to better illustrate the correspondence between the
$(R_{\tau,\sigma, i},
L_{\tau,\sigma, i})$
maps,
allocated in the target MPI process,
$\tau$, and the
$S_{\sigma, \tau, i}$
sequences, allocated in the source MPI process, $\sigma$.
Indeed, it can be observed that
the remote source neuron indexes in the map for
the source process 0
allocated in the target
process 1 are equal 
to the local source neuron indexes of the sequence
$S_{\sigma, \tau, i}$
allocated in the source process
0 for the target process 1,
in agreement with Eq. \ref{eq:mapalign}.
In this example, we assume that at a certain time step the neurons of index $480$ and $742$ of the MPI process of rank $0$, which are highlighted in yellow in the figure, emit a spike.

\begin{figure}[H]
    \centering
    \includegraphics[width=1\linewidth]{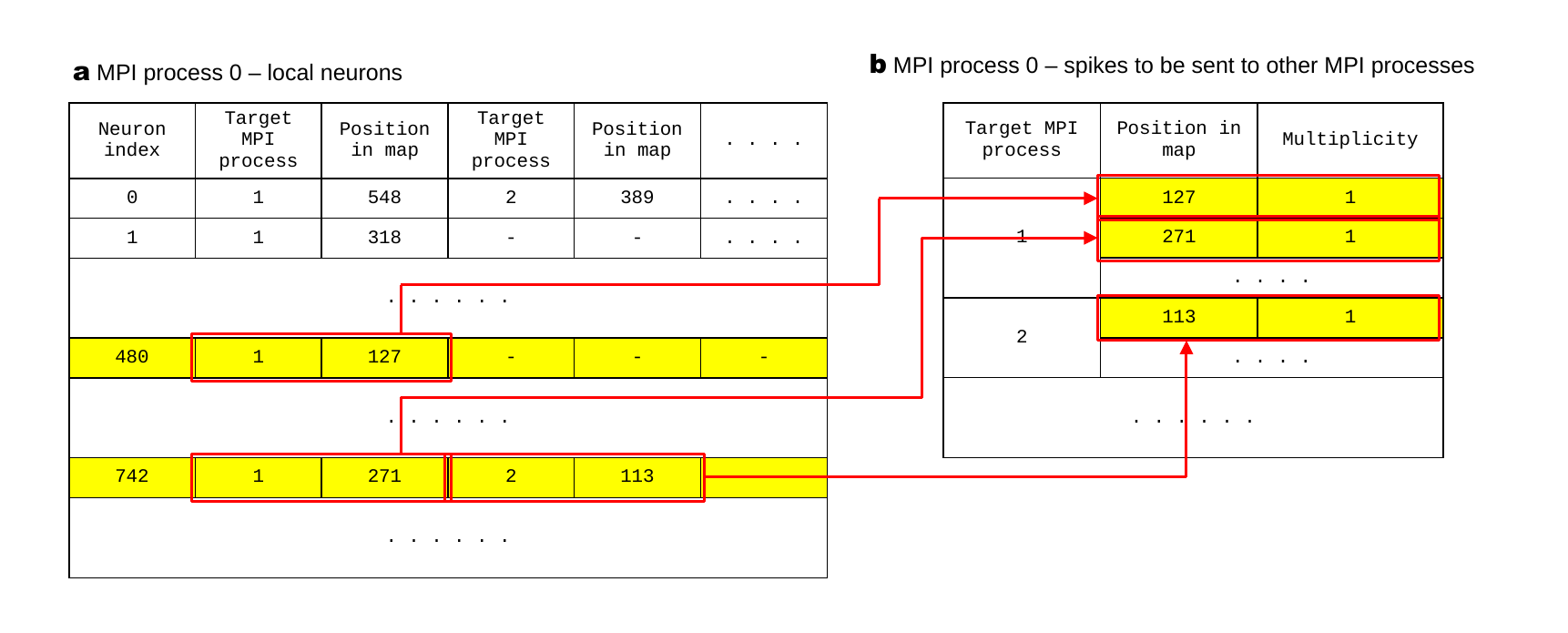}
    \caption{
(\textbf{a})
organization of the sequences
$T_{\sigma, s, j}$, 
which represent
the ranks of the target MPI processes in which 
each node has an image,
and $P_{\sigma, s, j}$
which are the corresponding
positions of the neuron in the maps that associate it to its image neurons in the target processes.
(\textbf{b})
creation of the
packets to be sent to target MPI processes using point-to-point MPI messages.
In this example we assume that in a certain time step nodes
480 and 742 of the MPI process 0
emit a spike. 
We can observe that node 480 has
an image in the target process 1, and that the corresponding position in the maps is 127,
while node 742 has images 
in the MPI processes 1 and 2,
and the corresponding positions 
are 271 and 113, respectively.
These positions are appended
to the packets created for the respective target MPI processes, which will be sent via point-to-point communication.
    }
    \label{fig:p2p-TP}
\end{figure}

Figure \ref{fig:p2p-TP}
\textbf{a}
shows how the sequences
$T_{\sigma, s, j}$ and
$P_{\sigma, s, j}$
are organized in 
the GPU memory.
The first column
contains the source neuron index, $s$, while the other columns
represent alternatively the values of the different target MPI processes
$T_{\sigma, s, j}$ in which the node $s$ has an image,
and the positions
$i \gets P_{\sigma, s, j}$ of the neuron in the maps
$(R_{\tau, \sigma, i}, L_{\tau, \sigma, i})$
that associate the source neuron to its image neurons in the target processes.

\begin{figure}[H]
    \centering
    \includegraphics[width=1\linewidth]{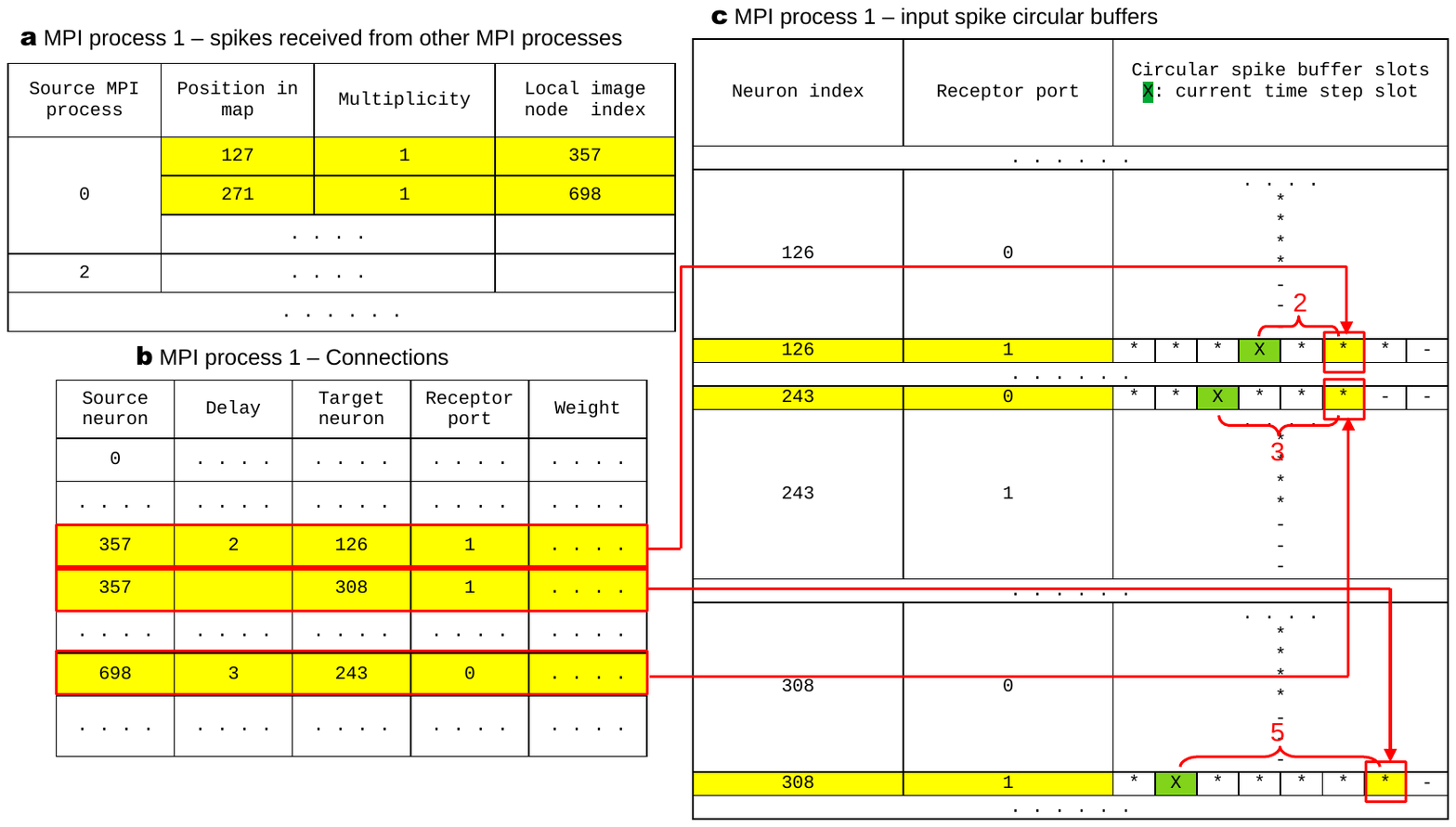}
    \caption{
Delivery of the spikes received by the target MPI processes.
(\textbf{a}) The first column indicates the source MPI process, the second column
represents the position
$i$ (127 and 271 in this example)
of the remote source neurons
that emitted the spike
in the map
$(R_{\tau, \sigma, i}, L_{\tau, \sigma, i})$
that associate the source neuron to its image neurons in the target processes,
the third column is the spike multiplicity, and the fourth
column contains
the values of
$L_{\tau, \sigma, i}$, which represents the local image neuron indexes.
(\textbf{b})
Connection structures allocated in the MPI process 1, sorted according to the source neuron 
index~\cite{Golosio23_9598}.
The connections outgoing
from local images of neurons
that emitted a spike
are highlighted in yellow.
In this example, the node 357 has two outgoing connections,
directed to the target neurons
126 and 308, with delays
2 and 5 in time-step units,
respectively,
while the node 698 has one
outgoing connection,
directed to the target neuron
243, having delay 3 in time-step units.
(\textbf{c}) The spikes are inserted into the input circular
spike buffers of the target neurons, in a slot shifted from the one representing the current time step by a number of positions equal to the delay
(2, 3 and 5 in the example, written in red). The spikes are accumulated in these slots by adding, for each of them, its multiplicity multiplied by the connection weight.
    }
    \label{fig:p2p-delivery}
\end{figure}

Figure~\ref{fig:p2p-TP}
\textbf{b}
illustrates how
 the spikes are inserted
 in packets that are sent to the target processes 
 through point-to-point MPI
 communications.
Finally,
figure \ref{fig:p2p-delivery}
shows how the spikes
received by the target MPI
process are delivered
to the target neurons
using the maps
$(R_{\tau, \sigma, i}, L_{\tau, \sigma, i})$, through the connection structures and the input spike circular buffers of the target neurons.

\end{document}